\documentclass[a4paper,11pt]{article}
\pdfoutput=1 % if your are submitting a pdflatex (i.e. if you have
\usepackage{jheppub}
\usepackage{dirtytalk}% for details on the use of the package, please
\usepackage[T1]{fontenc} % if needed
\title{\boldmath Superconducting multi-vortices and a novel BPS bound in chiral perturbation theory}

\author[a,b]{Fabrizio Canfora}
\author[c]{Marcela Lagos,}
\author[d,e]{Aldo Vera}

\affiliation[a]{Universidad San Sebasti\'{a}n, sede Valdivia, General Lagos 1163, Valdivia 5110693, Chile}
\affiliation[b]{Centro de Estudios Cient\'{\i}ficos (CECS), Casilla 1469, Valdivia, Chile}
\affiliation[c]{Universidad San Sebastián, Avenida del Cóndor 720, Santiago, Chile}
\affiliation[d]{N\'ucleo Matem\'aticas F\'isica y Estad\'istica, Universidad Mayor, Avenida Manuel Montt 367, Santiago, Chile.}
\affiliation[e]{Centro Multidisciplinario de F\'isica, Vicerrector\'ia de Investigaci\'on, Universidad Mayor, Camino La Pir\'amide 5750, Santiago, Chile.}

\emailAdd{fabrizio.canfora@uss.cl}
\emailAdd{marcela.lagos@uss.cl}
\emailAdd{aldo.vera@umayor.cl}

\abstract{We derive a novel BPS bound from chiral perturbation theory minimally
coupled to electrodynamics at finite isospin chemical potential. At a
critical value of the isospin chemical potential, a system of three
first-order differential field equations (which implies the second-order
field equations) for the gauge field and the hadronic profile can be derived
from the requirement to saturate the bound. These BPS configurations
represent magnetic multi-vortices with quantized flux supported by a
superconducting current. The corresponding topological charge density is
related to the magnetic flux density, but is screened by the hadronic
profile. Such a screening effect allows to derive the maximal value of the
magnetic field generated by these BPS magnetic vortices, being $B_{\text{max}%
}=2,04 \times 10^{14} \, \text{G}$. The solution for a single BPS vortex is
discussed in detail, and some physical consequences, together with the
comparison with the magnetic vortices in the Ginzburg-Landau theory at
critical coupling, are described.}

\makeatletter
\gdef\@fpheader{}
\makeatother

\begin{document} 
\maketitle
\flushbottom

%%%%%%%%%%%%%%%%%%%%%%%%%%%%%%%%%%%%%%%%%%%%
\section{Introduction}  \label{sec-1}
%%%%%%%%%%%%%%%%%%%%%%%%%%%%%%%%%%%%%%%%%%%%

The phase diagram of the low energy limit of quantum chromodynamics (QCD)
under non-trivial external conditions (such as finite density, low temperatures and in
presence of strong magnetic fields) is a very hard nut to crack (see \cite{R0}-\cite{Pisarski1}, and references therein). Not
only analytic perturbative methods fail, but also lattice QCD may experience
some problems (see, for instance, Refs. \cite{sign1}-\cite{QGBook}).

An extremely important region of the phase diagram corresponds to low
temperatures, vanishing baryonic chemical potential ($\mu _{B}=0$) and
finite isospin chemical potential ($\mu _{I}\neq 0$) \cite{IsospinR1}-\cite%
{Brandt}. There are many reasons behind the relevance of this sector. First,
when $\mu _{I}\neq 0$ but $\mu _{B}=0$, there is no sign problem, so that
detailed results on the phase diagram are available in this region. Second,
in the seminal paper by Son and Stephanov \cite{IsospinR1}, it has been
shown that a transition occurs from the
hadronic phase to a superfluid phase and, eventually, to a superconducting
phase. It is expected that at low isospin density the ground state is a pion
condensate \cite{IsospinR1}, \cite{Massimo2}, \cite{Massimo3}, while a Fermi
liquid with Cooper pairing should appear at higher density. These results
have been confirmed by lattice studies and still offer many surprises,
as we will see.

In particular, the analysis of the interplay between strong interactions and
electromagnetic fields exhibit difficult unsolved problems at low
energies and temperatures. At these scales, the main role is played by the
topological solitons of QCD \cite{[4]}-\cite{WeinbergBook}. Topological solitons are characterized by a non-vanishing
topological charge, which prevents such classical configurations from
decaying into the trivial vacuum. Especially taking into account the Cooper
pairing appearing at finite $\mu _{I}$, it is clear that the most natural
topological solitons that should appear are superconducting vortices with
quantized magnetic flux. Indeed, vortex-like solitons are expected both in
superfluids and in superconductors. At finite density, many vortex-like
solutions have been constructed numerically, and many of the following
references have been important sources of inspiration as far as the present
work is concerned; \cite{Adhikari1}-\cite{Nitta4}.

Analytical results have also been obtained in Refs. \cite{US1}-\cite{US6}. However, these analytical
solutions (which describe inhomogeneous baryonic condensates with the shapes
of tubes and layers) have been obtained at finite baryon density but not at
finite isospin density\footnote{%
While, in the present case, we need vanishing baryonic density but
non-vanishing isospin density.} and, moreover, do not possess the required
quantized magnetic flux characteristic of vortices in superconductors. One
of the most important open theoretical questions in this area appears when
we consider the situation in the Ginzburg-Landau (GL) theory. In such a
theory, depending on the value of the Higgs coupling, one can be in a type-I
or in a type-II superconductor and, exactly at the transition, a remarkable
phenomenon appears. One can saturate a Bogomol'nyi–Prasad–Sommerfield bound (BPS bound in what follows) for the free energy per
unit of length of the vortices, where the magnetic flux plays the role of the
topological charge, and multi-vortices solutions (where the vortices can be
placed at arbitrary locations) appear \cite{Weinberg}.

Since $\mu _{I}$ is responsible for the Cooper pairing \cite{IsospinR1} (and
so $\mu _{I}$ plays a similar role to the Higgs coupling), one would expect
that at a special value of $\mu _{I}$ it should be possible to saturate a
suitable BPS bound (where the magnetic flux is expected to be the
corresponding topological charge) in such a way that multi-vortices
solutions should appear. Such a special value of $\mu _{I}$ should be a
signal of a change of behavior from type-I to type-II (in which case flux
tubes can penetrate the superconductor). Despite all the efforts, such a BPS
completion has not been found. In this paper, we will fill this gap.

We will work with chiral perturbation theory ($\chi$PT), which describes the
low energy limit of QCD; see \cite{CPT1}-\cite{Massimo1}, and references therein.
As it will be clarified in the following sections, the present novel results
are well within the range of validity of $\chi $PT. Hence, our starting
point is the non-linear sigma model (NLSM) coupled to the Maxwell theory in
the case of $SU(2)$ isospin global symmetry, which is one of the most
relevant effective field theories \cite{[4]}, \cite{BaMa}. The analysis of
soliton crystals in $\chi $PT in the presence of magnetic fields and/or at
finite isospin chemical potential is a very active topic (see \cite{Schmitt2}-\cite{Nitta4},
and references therein). The conditions allowing the appearance of
topologically stable non-homogeneous condensates, besides its intrinsic
interest, are also relevant for the thermodynamics of dense nuclear/quark
matter \cite{R0}-\cite{R2}. Moreover, when the $SU(2)$ valued
scalar field is assumed to be homogeneous, and the corresponding spatial
fluctuations are neglected, many important physical effects are missed (see 
\cite{ex4d6}, \cite{ex4d4}, and references therein).

Until very recently, the more common techniques (which allow describing
stable topological solitons such as instantons, monopoles and vortices in
Yang-Mills-Higgs and the abelian-Higgs models) did not work in the case of
the gauged $\chi $PT. The reason is that, at the level of the gauged NLSM,
there is no obvious BPS bound to saturate for static configurations. Indeed,
the obvious bounds (where one would put, on the right-hand side, the
baryonic charge density or the magnetic flux density) cannot be saturated.
On the other hand, the recent results in Ref.  \cite{US7} have shown that when the
obvious BPS bound cannot be saturated, a novel less obvious BPS bound exist
that can actually be saturated, and it has all the expected properties
that a BPS bound should have. In the present paper, we will see a quite
remarkable manifestation of this phenomenon; \textit{we will find a novel
BPS bounds for a special value of $\mu _{I}$ in the gauged $\chi $PT which
can actually be saturated by superconducting multi-vortices with quantized
magnetic flux}.

The reason why this bound has not been found before is very tricky. The
topological charge density, which is needed for the BPS completion, is not
the magnetic flux density (as it happens in the GL case), but rather a
magnetic flux density dressed by the hadronic profile itself. Such a
dressing screens the magnetic field $B_{z}$ providing with a maximal
possible value for $B_{z}$. It is worth emphasizing
that magnetized topological solitons are extremely relevant in many
situations, such as heavy ions collisions and neutron stars (see Refs. \cite{sign1}-\cite{QGBook}, and \cite{strongB1}-\cite{strongB3}). A further by-product of the present results is that the Landau-Peierls criteria is avoided.

The techniques introduced in Refs. \cite{US1}-\cite{US6}, allowed the construction of analytical
solutions describing multi-solitons at finite density for both in the NLSM
as well as in the Skyrme model minimally coupled to the Maxwell theory (see,
for details, Refs. \cite{US8}-\cite{US14}). However, in such an approach, the electromagnetic
field (although self-consistently generated by the non-homogeneous
condensates) has necessarily the electric components of the same order as
the magnetic components. Therefore, in all the situations in which the
magnetic field dominates, this framework must be modified. This is the aim
of the present paper, in which the technique introduced in Ref. \cite{US7} will
be used to derive a novel BPS bound that can be saturated for magnetized
configurations.

This paper is organized as follows: In Section \ref{sec-2} the gauged $\chi$%
PT will be introduced together with the Ansatz for the matter fields. In
Section \ref{sec-3} we derive a novel BPS bound from the gauged $\chi $PT.
In Section \ref{sec-4} we construct magnetized vortex solutions from the BPS
equations and we discuss its main physical properties. In Section \ref{sec-5}
we show in detail the case of a single vortex. Section \ref{sec-6} is
devoted to the conclusions. In the Appendix we will provide with
mathematical arguments supporting the existence of such BPS multi-vortices
configurations.

%%%%%%%%%%%%%%%%%%%%%%%%%%%%%%%%%%%%%%%%%%%%
\section{Preliminaries} \label{sec-2}
%%%%%%%%%%%%%%%%%%%%%%%%%%%%%%%%%%%%%%%%%%%%

The gauged $\chi$PT (up to order $\mathcal{O}(p^{2})$) \cite{4m}, is
described by the action 
\begin{gather}  \label{I}
I[U, A]=\frac{1}{4e^2}\int_{\mathcal{M}} d^{4} x \sqrt{-g}\left(K \text{Tr}%
\left[L^{\mu} L_{\mu}\right] - F_{\mu \nu} F^{\mu \nu} \right) \ , \\
L_{\mu}=U^{-1} D_{\mu} U= L_{\mu}^{j} t_{j} \ , \quad F_{\mu\nu}=
\partial_\mu A_\nu - \partial_\nu A_\mu \ , \quad t_{j}=i \sigma_{j} \ . 
\notag
\end{gather}
Here $U(x)\in SU(2)$ is the pionic field, $A_{\mu }$ is the Maxwell
potential and $\sigma _{i}$ are the Pauli matrices. The coupling $K$ is a
positive constant given by $K=(\frac{ef_{\pi }}{2})^{2}$, being $f_{\pi }$
the pions decay constant and $e$ the electric charge.
Here we will use $f_{\pi }=93\,\text{MeV}$ and $e=\sqrt{4\pi \alpha }=0.303$. In our convention $c=1$ and $\hslash =1$. In Eq. \eqref{I}, $D_{\mu }$
denotes the covariant derivative, defined as 
\begin{equation}
D_{\mu }U=\nabla _{\mu }U+A_{\mu }U\hat{O}\ ,\quad \hat{O}=U^{-1}\left[
t_{3},U\right] \ ,  \label{covdev}
\end{equation}%
being $\nabla _{\mu }$ the partial derivative.\footnote{Note that in some references a normalization factor $1/2$ is used in front of $A_{\mu}$ in Eq. \eqref{covdev}.} The isospin chemical
potential can be introduced to the model through the covariant derivative in
the following form 
\begin{equation}
D_{\mu }U\rightarrow \bar{D}_{\mu }U=D_{\mu }U+\mu _{I}[t_{3},U]g_{\mu t}\ ,
\label{muI}
\end{equation}%
where $\mu _{I}$ is the value of isospin chemical potential. It is important
to point out that critical values for the isospin chemical potential have
been derived from effective models (such as the Skyrme model or ladder-QCD) showing
different phase transitions (see, for instance, Refs. \cite{Barducci}-\cite{Cao}). In the next section, we will compute a
critical value for $\mu _{I}$ in $\chi $PT that allows the formation of a
topologically stable inhomogeneous condensate made of magnetized multi-vortices.\footnote{%
In this section we will not include the pions mass since this chiral limit
illustrate the emergence of this phenomenon in a very neat way. The
inclusion of the pions mass will be discussed in Section \ref{sec-4}.}

The electromagnetic current obtained from Eq. \eqref{I} is given by 
\begin{equation}  \label{J}
J_\mu = - \frac{K}{2} \text{Tr} (\hat{O} L_\mu) \ .
\end{equation}
The pionic field in the exponential representation is written as 
\begin{gather}
U=\cos (\alpha )\boldsymbol{1}+\sin (\alpha )n_{i}t^{i}\ ,  \label{U} \\
n_{i}=\{\sin \Theta \cos \Phi ,\sin \Theta \sin \Phi ,\cos \Theta \}\ , 
\notag
\end{gather}%
where $\alpha =\alpha (x^{\mu })$, $\Theta =\Theta (x^{\mu })$, $\Phi
=\Phi(x^{\mu })$ are the three degrees of freedom of the $U$ field, and $%
\boldsymbol{1}$ denotes the $2\times2$ identity matrix. In terms of this
parametrization, the covariant derivative $D_{\mu }U$ in Eq. (\ref{covdev})
reads%
\begin{equation}  \label{covdev2}
D_{\mu }\alpha = \partial _{\mu }\alpha \ , \quad D_{\mu }\Theta
=\partial_{\mu }\Theta \ , \quad \ D_{\mu }\Phi =\partial _{\mu }\Phi -2
A_{\mu }\ .
\end{equation}
Hence, one can see that the scalar degree of freedom $\Phi (x^{\mu })$ plays
the role of the phase of the complex Higgs field in the GL theory (while $%
\alpha $ and $\Theta $\ are not affected by the covariant derivative).

Since we are interested in multi-vortices (with quantized magnetic field
along the third spatial direction) in $\chi $PT minimally coupled to the
Maxwell theory, the natural Ansatz is 
\begin{gather}
\alpha =\alpha (x_{1},x_{2})\ ,\quad \Phi =\Phi (x_{1},x_{2})\ ,\quad \Theta
=\frac{\pi }{2}\ ,  \label{U1} \\
A_{\mu }dx^{\mu }=A_{1}\,dx_{1}+A_{2}\,dx_{2}\ ,\qquad
A_{i}=A_{i}(x_{1},x_{2})\ ,  \label{Amu}
\end{gather}%
which leads to a magnetic field in the $x_{3}$ direction. This choice is
very convenient since, as it will be clear from the following computations,
the combination $\sin (\alpha )n_{i}t^{i}$, plays the role of the complex
Higgs field $\phi $\ in the GL free energy: 
\begin{equation*}
\sin (\alpha )n_{i}t^{i}\backsim \phi \ ,
\end{equation*}%
while the choice of the gauge potential is forced by the requirement to have
a magnetic field along $x_{3}$.

%%%%%%%%%%%%%%%%%%%%%%%%%%%%%%%%%%%%%%%%%%%%
\section{A novel BPS bound from $\chi$PT} \label{sec-3}
%%%%%%%%%%%%%%%%%%%%%%%%%%%%%%%%%%%%%%%%%%%%

In this section, for simplicity, we will assume a flat metric in Cartesian
coordinates, but our results can be obtained in the same way using arbitrary
coordinate systems, as we will see in the next section.

When one replaces in Eq. \eqref{I} the Ansatz defined in Eqs. \eqref{U}, %
\eqref{U1} and \eqref{Amu}, (taking into account the non-trivial isospin
chemical potential in Eq. \eqref{muI}) one obtains the lagrangian density $%
\mathcal{L}$, which will play the role of free energy density $\digamma $,
that is 
\begin{equation}
\digamma= -\mathcal{L} = \frac{1}{e^2}\biggl[ \frac{K}{2}\biggl\{\left( \overrightarrow{\nabla }%
\alpha \right) ^{2}+\sin ^{2}(\alpha )\left( \overrightarrow{D}\Phi \right)
^{2}\biggl\}+\frac{1}{2}(\vec{B}^{2})-2K\mu _{I}^{2}\sin ^{2}(\alpha )\ \biggl] \ ,
\label{L}
\end{equation}
where we have denoted 
\begin{equation*}
\left( \overrightarrow{\nabla }\alpha \right) ^{2} = \sum_{i=1}^{2}\left(
\partial _{i}\alpha \right) ^{2}\ ,\quad \left( \overrightarrow{D}\Phi
\right) ^{2}=\sum_{i=1}^{2}\left( D_{i}\Phi \right) ^{2} \ .
\end{equation*}
In order to derive a BPS bound from Eq. \eqref{L}, it is convenient to
rewrite the free energy density as follows: 
\begin{equation}
e^2 \digamma =\frac{K}{2}\biggl\{\left( \overrightarrow{\nabla }\alpha \right)
^{2}+\sin ^{2}(\alpha )\left( \overrightarrow{D}\Phi \right) ^{2}\biggl\}+%
\frac{1}{2}(\vec{B}^{2})+2K\mu _{I}^{2}\cos ^{2}(\alpha) -2K\mu _{I}^{2}\ .
\label{F1}
\end{equation}%
One can see that the above expression would be positive definite if one
could eliminate the last constant term. From the viewpoint of the
differential equations, the constant term is certainly not important (and
one could simply drop it). However, such a term is telling us something
physical: in the present case, the appropriate thermodynamic potential is
not the free energy $\boldsymbol{F}$, but rather the Gibbs free energy $%
\boldsymbol{G}$ 
\begin{equation}
\boldsymbol{G}=\boldsymbol{F}+PV\ , \qquad \boldsymbol{F}=\int d^{3}x \, \digamma \ ,
\quad \boldsymbol{G} = \int d^{3}x \, G \ ,  \label{Gibbs0}
\end{equation}%
being $V$ the volume and $P$ the pressure fixed by the chemical potential,
namely $P=2K\mu _{I}^{2}$, according to Eq. \eqref{F1}. Thus, the Gibbs free energy density $G$ reads%
\begin{equation}
e^2 G =\frac{K}{2}\biggl\{\left( \overrightarrow{\nabla }\alpha \right)
^{2}+\sin ^{2}(\alpha )\left( \overrightarrow{D}\Phi \right) ^{2}\biggl\}+%
\frac{1}{2}(\vec{B}^{2})+2K\mu _{I}^{2}\cos ^{2}(\alpha) \ ,  \label{Gibbs1}
\end{equation}
so that 
\begin{equation}  \label{Gibbs2}
\frac{\boldsymbol{G}}{L}\boldsymbol{=}\int dx_{1}dx_{2}G \ .
\end{equation}
Eq. \eqref{Gibbs2} has to do with the fact that we are interested in
superconducting vortices whose Gibbs free energy does not depend on the
third direction. From now on, we will work with $G$ since it is a positive
definite quantity, and it simplifies the discussion of the boundary
conditions.

\subsection{Boundary conditions and flux quantization}

The requirement to have a finite Gibbs free energy per unit of length gives
rise to boundary conditions that are close analogs of the boundary
conditions used for vortices in the GL theory.

Let us introduce the usual coordinate system to discuss boundary conditions
for a single vortex at the origin of the cylindrical coordinates:
\begin{equation*}
ds^{2}=-dt^{2}+dr^{2}+r^{2}d\theta ^{2}+\left( dx_{3}\right) ^{2}\ .
\end{equation*}%
In order for the integral in Eq. \eqref{Gibbs2} to be convergent for large
values of the two-dimensional radius, we must require 
\begin{equation}
\alpha \underset{r\rightarrow \infty }{\rightarrow }\frac{\pi }{2}+c_1\pi \ ,\
c_1\in \mathbb{N}\ ,\qquad \qquad \overrightarrow{D}\Phi \underset{%
r\rightarrow \infty }{\rightarrow }0\ \Leftrightarrow \overrightarrow{A}\ 
\underset{r\rightarrow \infty }{\rightarrow } \frac{1}{2}\overrightarrow  {\partial }\Phi
\ .  \label{bc1}
\end{equation}%
On the other hand, close to the origin, regularity requires%
\begin{equation}
\alpha \underset{r\rightarrow 0}{\rightarrow }c_2\pi \ ,\ c_2\in \mathbb{N}\
,\qquad \qquad \overrightarrow{A}\ \underset{r\rightarrow 0}{\rightarrow }0\
.  \label{bc2}
\end{equation}%
The condition in Eq. \eqref{bc2}, from the viewpoint of the chiral field $U$
in Eqs. \eqref{U} and \eqref{U1}, reads: 
\begin{equation}
U\underset{r\rightarrow 0}{\rightarrow }\pm \boldsymbol{1}\ .  \label{bc3}
\end{equation}%
In fact, Eqs. \eqref{bc2} and \eqref{bc3} must be satisfied at the locations
of all the vortices as one can always choose a local coordinates system
where the position of the $j-$th vortex is the origin: 
\begin{equation}
U\underset{r\rightarrow r_{j}}{\rightarrow }\pm \boldsymbol{1}\ ,\quad
\forall \ j=1,..,N\ ,  \label{bc6}
\end{equation}%
where $N$ is the total number of vortices and $r_{j}$ is the position of the 
$j-$th vortex.

On the other hand, one can derive the quantization of the magnetic flux in
the usual way. The chiral field $U$ in Eq. \eqref{U} must be single valued:%
\begin{equation}
U(r,\theta )=U(r,\theta +2\pi )\ \ ,  \label{singlevalued}
\end{equation}%
and then, taking into account Eq. \eqref{U1}, one arrives at the conclusion
that 
\begin{equation}
\Phi (r,\theta )=\Phi (r,\theta +2\pi )+2n \pi \ ,\ \ n\in \mathbb{N}\ ,
\label{multivalued}
\end{equation}%
which is natural taking into account that $\Phi $ plays the role of the
phase of the complex Higgs field in the GL theory (see Eq. \eqref{covdev2}).
This implies that the following integral is an integer 
\begin{equation}
\frac{1}{2\pi } \oint_{\Gamma _{\infty }} \overrightarrow{\partial }%
\Phi \cdot d\overrightarrow{l}=n\ ,  \label{fluxquantization}
\end{equation}
where $\Gamma _{\infty }$ is the circle at spatial infinity. It is worth
emphasizing here that the above integral would be an integer for any closed
curve $\Gamma $ enclosing all the points $\left\{ r_{j}\right\} _{j=1,..,N}$%
, where $\sin(\alpha) $ vanishes. On the other hand, the finiteness of the
Gibbs free energy requires Eq. \eqref{bc1}. Consequently, due to the Stokes
theorem, as in the case of the GL theory, one obtains that $n$ is
exactly the magnetic flux, and it is quantized.

\subsection{Improper gauge transformations}

In order to derive a BPS bound, we rewrite Eq. \eqref{Gibbs1} as 
\begin{align*}
e^2 G=& \frac{K}{2}\biggl((\partial _{1}\alpha )^{2}+(\partial _{2}\alpha
)^{2}+\sin ^{2}(\alpha )[(D_{1}\Phi )^{2}+(D_{2}\Phi )^{2}])\biggl)+\frac{1}{%
2}B_{z}^{2}+2K\mu _{I}^{2}\cos ^{2}(\alpha ) \\
=& \frac{K}{2}\biggl((\partial _{1}\alpha +\sin (\alpha )D_{2}\Phi
)^{2}+(\partial _{2}\alpha -\sin (\alpha )D_{1}\Phi )^{2}\biggl)+\frac{1}{2}%
\biggl(B_{z}-2\mu _{I}\sqrt{K}\cos (\alpha )\biggl)^{2} \\
& -K\sin (\alpha )(\partial _{1}\alpha D_{2}\Phi -\partial _{2}\alpha
D_{1}\Phi )+2\mu _{I}\sqrt{K}\cos (\alpha )B_{z}\ .
\end{align*}%
Note that, \textit{locally}, we can gauge away the phase $\Phi $ as follows 
\begin{gather}
A_{\mu }\rightarrow \widehat{A}_{\mu }=A_{\mu }+ \frac{1}{2}\partial _{\mu }\Phi \ ,
\label{gt1} \\
\Phi \rightarrow \widehat{\Phi }=\Phi -\Phi =0\ .  \label{gt2}
\end{gather}%
However, it is very important to emphasize that $\Phi $ itself is a
multi-valued function precisely due to the non-vanishing magnetic flux (see
Eqs. \eqref{multivalued} and \eqref{fluxquantization}). Therefore, Eqs. %
\eqref{gt1} and \eqref{gt2} do not define a proper gauge transformation. In
order to clarify which is the price to be paid (in particular, which extra
terms do appear) when one performs a gauge transformation with a
multi-valued gauge parameter, let us consider the case of a single vortex
(which will be analyzed in more detail in Section \ref{sec-5}). The
corresponding Ansatz is 
\begin{equation}
\alpha =\alpha (r)\ ,\quad A_{r}=A_{z}=0\ ,\quad A_{\theta }=A(r)\ ,\quad
\Phi =n\theta \ ,  \label{singlevortex1}
\end{equation}%
where the cylindrical coordinates are 
\begin{equation}
ds^{2}=-dt^{2}+dr^{2}+r^{2}d\theta ^{2}+dz^{2}\ .  \label{singlevortex2}
\end{equation}%
The spatial coordinates $\{r,\theta ,z\}$, have the ranges 
\begin{equation*}
0<r<\infty \ ,\quad 0\leq \theta <2\pi \ ,\quad -\infty <z<\infty \ .
\end{equation*}%
As one can see from Eq. \eqref{covdev2}, the interaction between the matter
field and the gauge field appears through the combination $\partial _{\mu
}\Phi -2A_{\mu }$. It is clear that with the Ansatz in Eq. %
\eqref{singlevortex1}, the conditions of regularity at the origin in Eq. %
\eqref{bc2} reduce to%
\begin{equation*}
A_{\theta }=A(r)\underset{r\rightarrow 0}{\rightarrow }0\ ,\qquad \sin
\alpha (r)\underset{r\rightarrow 0}{\rightarrow }0\ ,
\end{equation*}%
which means, of course, that the origin is the position of the vortex. If
one now performs the improper gauge transformation in Eqs. \eqref{gt1} and %
\eqref{gt2} one would get 
\begin{equation*}
\widehat{A}_{\theta }=A(r)+ \frac{1}{2} n\underset{r\rightarrow 0}{\rightarrow } \frac{1}{2} n\
,\qquad \sin \alpha (r)\underset{r\rightarrow 0}{\rightarrow }0\ .
\end{equation*}%
This is not an innocuous shift of the gauge potential since it introduces a
singularity. This is most easily seen in Cartesian coordinates, where%
\begin{equation*}
\Phi =n\theta =n\arctan \left( \frac{y}{x}\right) \ .
\end{equation*}%
A gauge potential $A_{\mu }$ proportional to $\partial _{\mu }\Phi $, with
the $\Phi $ here above, is nothing but the gauge potential of an
Aharonov-Bohm gauge field. As it is well known, such a gauge potential,
despite being locally a pure gauge, it corresponds to the magnetic field of a
very thin solenoid along the $z$ axis. Therefore, the magnetic field, $%
B_{z}^{AB}$, associated with the Aharonov-Bohm gauge field is%
\begin{equation}
B_{z}^{AB}=n\delta (x,y)\ ,  \label{AB-flux}
\end{equation}%
where $\delta (x,y)$ is the Dirac delta at the position of the thin
solenoid. This argument is true actually at the positions of all the
vortices. The reason is that, close to the position of any of the vortices
present in the configuration, one can choose a coordinate system with the
origin at the position of such vortex and repeat this argument.
Consequently, the price to pay if one decides to \textquotedblleft gauge
away" the function $\Phi $ (which is convenient from the computational point
of view) in a configuration with $N$ vortices is that one has to include
explicitly in the magnetic field the \textquotedblleft Aharonov-Bohm
contributions" from all the vortices: 
\begin{equation}
B_{z}^{AB}=\sum_{j=1}^{N}n_{j}\delta \left( x-x_{j},y-y_{j}\right) \ ,
\label{AB-flux2}
\end{equation}%
(note that we have absorbed the factors of $4\pi $ in the $n_{j}$ in order
to simplify the notation) which corresponds to a $\Phi $ of the following
form%
\begin{equation*}
\Phi =\sum_{j=1}^{N}n_{j}\arg \left( z-z_{j}\right) \ ,
\end{equation*}%
where $z=x+iy$, and $z_{j}$ is the position of the $j-$th vortex in complex
coordinates (compare, for instance, with the expression below Eq. (3.2) of
Ref. \cite{YsonnYang}). Of course, this same result has been obtained in a
more rigorous way in the classic references \cite{Taubes1} and \cite{Taubes2}
. In particular, the above expression corresponds to the typical sum over
the Dirac deltas, which appear in the mathematical literature (see, for
instance, the right hand side of Eq. (3.3) in Ref. \cite{YsonnYang}).
Although the present argument is less rigorous than the one in the above
mentioned references, it offers a very intuitive physical picture: if one
eliminates $\Phi $ with an improper gauge transformation, then the
\textquotedblleft Aharonov-Bohm contributions" from all the vortices in Eq. %
\eqref{AB-flux2} must be included.

\subsection{BPS completion}

The elimination of $\Phi $ simplifies the derivation of the BPS bound. One
can rewrite the free energy density as 
\begin{align}
e^2 G=& \frac{K}{2}\biggl((\partial _{1}\alpha +\sin (\alpha) \widehat{A}%
_{2})^{2}+(\partial _{2}\alpha -\sin (\alpha) \widehat{A}_{1})^{2}\biggl)+%
\frac{1}{2}\biggl(\widehat{B}_{z}-2\mu _{I}\sqrt{K}\cos (\alpha )\biggl)^{2}
\label{G22} \\
& -K\sin \alpha (\partial _{1}\alpha \widehat{A}_{2}-\partial _{2}\alpha 
\widehat{A}_{1})+2\mu _{I}\sqrt{K}\cos (\alpha )\widehat{B}_{z}\ .  \notag
\end{align}
In order to obtain the sought BPS completion, we need to write the last two
terms in the above equation as a total derivative. This is only possible at
the following special critical value for the isospin chemical potential 
\begin{equation}
\mu _{I}^{\text{c}}= \sqrt{K} = \frac{ef_{\pi }}{2}=14,1 \,\text{MeV} \ .
\label{critical1}
\end{equation}%
Quite remarkably, this particular value is well within the range of
validity of $\chi$PT \cite{IsospinR7}. In fact, this value allows that 
\begin{gather} \notag 
-K\int dx^{1}dx^{2}\biggl[\sin (\alpha) (\partial _{1}\alpha \,\widehat{A}%
_{2}-\partial _{2}\alpha \,\widehat{A}_{1})-\cos (\alpha )\widehat{B}_{z}%
\biggl]=\int d\omega \ , \\ 
\omega =\cos (\alpha )(\widehat{A}_{1}dx^{1}+\widehat{A}_{2}dx^{2}) \ ,
\end{gather}%
where we have used that $\widehat{B}_{z}=F_{12}=\partial _{1}\widehat{A}%
_{2}-\partial_{2}\widehat{A}_{1}$. Thus, the Gibbs free energy density is
minimized when the following first order equations are satisfied 
\begin{align}
\partial _{1}\alpha +\sin (\alpha) \,\widehat{A}_{2}& =0\ ,  \label{BPS1} \\
\partial _{2}\alpha -\sin (\alpha) \,\widehat{A}_{1}& =0\ ,  \label{BPS2} \\
\widehat{B}_{z}-2K\cos (\alpha )& =0\ ,  \label{BPS3}
\end{align}%
and the following BPS bound emerges 
\begin{equation} \label{BOUND}
    \boldsymbol{G} \geq   \int_{\partial \Sigma}  d\omega  \ ,
\end{equation}
where $\partial \Sigma$ is usually taken as the $S^1$ circle at spatial infinity. 

It is important to note that the value of the free energy
depends on the topological invariant that is fixed by the boundary conditions on the sample. Thus, even when the pressure and chemical potential are fixed, the BPS bound allows the existence of configurations with different $\boldsymbol G$ determined by the choice of the boundary conditions. In particular, for the  BPS vortex phase and the homogeneous phase with the same $(B, P, \mu_I)$ and different free energies:
if the boundary conditions are such that the topological invariant
vanishes, then the homogeneous phase has lower free energy than the vortex
phase, while if the boundary conditions are such that the topological
invariant is non-trivial, then the vortex phase wins. This also happens for normal superconductors at a special critical coupling for Ginzburg-Landau theory, which allows BPS vortices.
One has, at a special point in the parameter space, both the homogeneous solution and the BPS crystal, and  which solution is favored is decided by the boundary conditions.

The relation between the topological charge on the right hand side of the above equation and the magnetic flux will be discussed in the following sections.

It is interesting to note that, for small $\alpha $, the above BPS system
reduces exactly to the BPS system of multi-vortices in critical
superconductors. In fact, considering 
\begin{equation*}
\left\vert \alpha \right\vert \ll 1\ ,\ \ \sin (\alpha )\sim \alpha \ ,\
\cos (\alpha )\sim 1-\frac{\alpha ^{2}}{2} \ ,
\end{equation*}
we obtain the well-known system 
\begin{align*}
\partial _{1}\alpha +\alpha \,\widehat{A}_{2}& =0\ , \\
\partial _{2}\alpha -\alpha \,\widehat{A}_{1}& =0\ , \\
\widehat{B}_{z}-K(1-\frac{\alpha ^{2}}{2})& =0\ .
\end{align*}%
This fact explains why the boundary conditions at the positions of the
vortices (where $\alpha $ is small) are the same as in the GL case.

In order to solve the BPS system in Eqs. \eqref{BPS1}, \eqref{BPS2} and %
\eqref{BPS3}, it is convenient to define the function $H(\alpha )=\ln {(\tan 
{(\alpha /2)})}$. It follows that 
\begin{equation*}
\widehat{A}_{1}=\frac{1}{\sin \alpha }\partial _{2}\alpha =\partial
_{2}H(\alpha )\ ,\quad \widehat{A}_{2}=-\frac{1}{\sin \alpha }\partial
_{1}\alpha =-\partial _{1}H(\alpha )\ .
\end{equation*}%
Using the above expression for the Maxwell potential, the first order
equations in Eqs. \eqref{BPS1}, \eqref{BPS2} and \eqref{BPS3}, can be
written as a second order differential equation for $H(\alpha )$, namely, 
\begin{equation}
\nabla ^{2}H=K\tanh (H)+\sum_{j=1}^{N}n_{j}\delta \left(
x-x_{j},y-y_{j}\right) \ ,  \label{MasterEquation}
\end{equation}%
where the second term on the right-hand side corresponds to the
``Aharonov-Bohm contributions" from all the vortices.\footnote{%
The comparison between the above non-linear equation determining BPS
superconducting vortices in $\chi $PT at finite isospin chemical potential
and the one characterizing superconducting vortices in the GL theory at
critical coupling (together with a sketch of a proof of the existence of
solutions of Eq. \eqref{MasterEquation} for any distribution of vortices)
will be delayed in the Appendix.} This equation (without the $\delta -$like
terms) can be obtained from the variation of the following effective action: 
\begin{equation}
\mathcal{L}_{\text{eff}}(H)=\frac{1}{2}(\vec{\nabla}H)^{2}+K\ln {(2\cosh H)}%
\ .  \label{effectiveaction1}
\end{equation}

\subsection{Advantages of the availability of BPS bounds}

The appearance of a BPS configuration arises
from a delicate balance between opposite forces (typically, when attractive
and repulsive interactions of the theory of interest are of the same order), one of the most important examples being BPS vortices in superconductors.
Despite the fact that these configurations appear at special points in
parameters space, they have fundamental importance for several reasons (see \cite{[4]}-\cite{WeinbergBook} and references therein). Here
we will mention just a few of them. 
\begin{enumerate}
    \item  These configurations signal a
transition from one type of behavior to a different type of behavior (as in the transition from type-I to type-II superconductors).
\item The
physical effects generated by BPS configurations are non-perturbative in
nature, namely that it cannot be captured by perturbation theory as the dependence of the relevant physical quantities on the coupling
constants is manifestly non-analytic. 
\item  There is
a very powerful tool to describe the low-energy dynamics of these BPS
configurations (which is not available for non-BPS configurations) in terms
of geodesics on moduli space. Such a tool allows for the description of dynamical
processes (such as vortex-vortex scattering), which are very difficult to
analyze with different methods. 
\item  BPS configurations are
topologically stable (if one starts from a positive definite action functional), then, these solutions cannot be destroyed by quantum/thermal fluctuations.

\item It is possible to derive many
exact results about the number of fermionic zero modes which BPS solitons
are able to support just looking at the topological charge: such results are
unavailable for generic configurations \cite{[5]}, \cite{WeinbergIndex}.
\end{enumerate}

These configurations are known to appear in supersymmetric theories as well
as in Yang-Mills-Higgs theory \cite{Marino} (where non-abelian monopoles, instantons and
so on are found in the non-perturbative sectors). On the other hand, to the best of author's knowledge, here we
constructed the first BPS vortices in $\chi$PT, which is
very far from being supersymmetric. It is also worth noticing that even in
the case of BPS vortices in BCS superconductors, the corresponding BPS
equations must be solved numerically (unlike what happens for instantons and
non-abelian monopoles). In fact, as it has already been stressed, the
importance of BPS vortices in superconductors lies in their existence and
not in the possibility of solving analytically the BPS equations.

At this point it is important to note relevant differences between the present results and the ones reported in Ref. \cite{Adhikari2}. In our case, in order to take full advantage of the BPS bound, we need to start from a free energy functional which is positive definite from the very beginning. Here, such a functional is the Gibbs free energy at the pressure in Eq. \eqref{Gibbs0}. Indeed, if we would use the free energy at zero external pressure, our BPS configuration would be saddle points but not necessarily minima in the corresponding topological sector. This means that the transition between type-I and type-II behavior, disclosed by the presence of the BPS vortices, appears at the value of the pressure given above. This is not surprising since external pressure affects heavily the properties of superconductors \cite{Roadmap}. On the other hand, the results in Ref.  \cite{Adhikari2} hold at zero external pressure.

Also, it is important to note that here $\boldsymbol{G}$ is simply the free energy with the inclusion of a
vacuum subtraction at fixed $\mu_I$, i.e., the free energy at vanishing
fields is subtracted. On the other hand, such a subtraction is not necessary if, instead of
the free energy one uses the Gibbs free energy, with the $PV$ term included.
Consequently, if one uses the Gibbs free energy instead of the free
energy, not only any subtraction is avoided, but also one gets directly a
positive definite functional, which ensures the property \textbf{4} in the
list here above. The interpretation of this fact is that
a region of type-I behavior opens up at this value of the
pressure. It is also
worth emphasizing that in our results there is no approximation, as we have
found the BPS point analytically, and due to the properties of BPS
configurations, our results are robust against perturbations (something that
cannot be said about generic vortices far from the BPS point). 

Regarding property \textbf{5}, let us shortly describe Weinberg’s computations in Ref. \cite{WeinbergIndex} to obtain the number of fermionic zero modes captured by our BPS vortices.

In order to compute the index, which is defined as
\begin{equation}
    \mathcal{I}= \text{dim}(\text{Ker}\ \mathcal{D})-\text{dim}(\text{Ker}\ \mathcal{D^\dagger}) \  ,
\end{equation}
one can follow Weinberg's technique, considering the limit of $M^2\rightarrow 0$ of the quantity
\begin{equation} \label{IM}
    \mathcal{I}(M^2)=\text{Tr}\biggl(\frac{M^2}{\mathcal{D}^\dagger\mathcal{D}+M^2}\biggl)-\text{Tr}\biggl(\frac{M^2}{\mathcal{D}\mathcal{D}^\dagger+M^2}\biggl) \ . 
\end{equation}
The expression in Eq. \eqref{IM} is extremely important because it counts the fermionic zero modes supported by the BPS configuration of interest. We would like to choose a suitable Dirac operator $\mathcal{D}$ in such a way that Weinberg's procedure applied to our case produces, as a result, the topological invariant appearing on the right-hand side of the BPS bound in Eq. \eqref{BOUND}. This reverse engineering process is worthless unless the resulting Dirac operator turns out to be physically interesting. The comparison with the Dirac equation analyzed in Eq. (3.3) of Ref. \cite{WeinbergIndex} suggests that we should take $\mathcal{D}$ as 
\begin{equation}
\mathcal{D}=\gamma^\mu(\partial_\mu-Q A_\mu)+U^{\gamma_5} \ ,\label{D}
\end{equation}
where $\gamma_5$ is needed for chirality reasons and $Q$ is the charge matrix of the quarks.
Indeed, in order to repeat the steps of Ref. \cite{WeinbergIndex} (see, in particular, Eqs. (3.1), (3.2) and (3.3) of Ref. \cite{WeinbergIndex}), the gauge potential and the chiral field $U$ should appear on the same footing (similarly as the gauge potential and the Higgs field appear in the case of vortices in BCS superconductors at critical coupling). On the other hand, $U^{\gamma_5}$ is necessary in the same way, as the Higgs field couples with fermions while its conjugate couples to anti-fermions. Quite interestingly, the above Dirac operator in Eq. \eqref{D} (whose form has been guessed with the \say{naive} requirement to reproduce the topological invariant in Eq. BPS upon using Weinberg's procedure) is compatible with the Dirac operator proposed in Refs. \cite{Jackiw:1981ee}-\cite{Kahana:1985ycl}, to describe the coupling of quarks with the chiral field $U$. The present arguments strongly suggest that the coupling  of quarks with the BPS vortices introduced in the previous section could share many properties with the coupling of electrons with BPS vortices in critical superconductors. The analysis of the coupling of quarks with the present BPS vortices will be studied in detail in a future publication.

%%%%%%%%%%%%%%%%%%%%%%%%%%%%%%%%%%%%%%%%%%%%
\section{Multi-vortices at critical values} \label{sec-4}
%%%%%%%%%%%%%%%%%%%%%%%%%%%%%%%%%%%%%%%%%%%%

The novel BPS bound presented in the previous section presents many peculiar
properties. In this section, we detail the most important ones.

\subsection{Strong magnetic field and finite isospin chemical potential}

\begin{enumerate}
\item First, the topological charge density, which appears naturally in
the BPS bound is not simply the magnetic flux density, as one would naively
expect; instead, it is a ``dressed magnetic flux" modulated by the hadronic
profile (this possibility has been pointed out for the first time in Ref. \cite%
{US7}). The presence of the pions generates a screening effect on the
magnetic flux, tending to suppress the magnetic field of the vortices in the
condensate.

\item The inhomogeneous condensate is stabilized by the topological
charge as it happens for the usual GL vortices, and, as in the GL theory,
the possibility to have multi-vortices configurations at arbitrary locations
appears at a special value of the relevant parameter (which is the Higgs
coupling in the GL case, while the isospin chemical potential in the
present case). From the viewpoint of differential equations, this could be
enough. However, from the physical viewpoint, the fact that these are
topologically non-trivial minima of the Gibbs free energy (and not of the
free energy) means that they can be realized in actual experiments keeping
the pressure fixed at the value $P= 2K (\mu_I^{\text{c}})^2= 858.474 \, (%
\text{MeV})^4$.

\item Unlike what happens in the GL theory, where the derivation of a
bound on the maximal magnetic field requires quite some work (see, for
instance, Refs. \cite{YsonnYang}-\cite{Taubes2}), the present
novel BPS bound allows to find very easily the maximum $B_{\max }$ value for
the magnetic field beyond which the condensate ceases to exist. Indeed,
looking at Eq. \eqref{BPS3}, one gets 
\begin{equation}
B_{\text{max}}=2K=2\biggl(\frac{ef_{\pi }}{2}\biggl)^{2}=397,03\,(\text{MeV}%
)^2 = 2,04 \times 10^{14} \, \text{G} \ .
\end{equation}

The above value for the maximal magnetic field is valid at the BPS point. Even more, it is still a good estimate for $B_{\text{max}}$ when the isospin chemical potential and pressure are close to corresponding values at the BPS bound. It is worth emphasizing that it is a novelty of our
approach to be able to compute exactly (thanks to the BPS properties of the
vortices) the maximal magnetic field even if just at a special point (as there are
very few exact results in the low energy limit of QCD).

\item The behavior of the electromagnetic field, the current and the
energy density can be easily described thanks to availability of the BPS
equations. First, as we have point out above, the magnetic field only
possesses component in the $z$ direction. This magnetic field is generated
by a self-sustained current, given by 
\begin{equation}
J_\mu = 2K \sin^2(\alpha) D_\mu \Phi \ .  \label{current}
\end{equation}%
From Eq. \eqref{current} one can see that the current is not-null even when
the electromagnetic field is suppressed. In fact, there is a persistent
current generated by the coupling with pions, given by $J_\mu^{(0)} =2K
\sin^2(\alpha) \partial_\mu \Phi$.

\item The stationary point of the free energy density $\digamma $ can
be determined through the variation of Eq. \eqref{F1} with respect to the
field $\alpha$, $\Phi$ and $A_\mu$, leading to the following set of second
order field equations  
\begin{gather}
\triangle \alpha - \sin(\alpha)\cos(\alpha)\biggl((\vec{D}\Phi)^2-4\mu_I^2%
\biggl) = 0 \ ,  \notag \\
\triangle \Phi - 2\ \vec\nabla\cdot \vec A + 2\cot(\alpha)\vec\nabla
\alpha\cdot\vec D\Phi = 0 \ ,  \label{secondorder} \\
\partial_j F^{ij} - 2 K \sin^2(\alpha) D^{i}\Phi = 0 \ ,  \notag
\end{gather}
where $\triangle$ is the Laplacian operator. As expected, the first order
BPS equations in Eqs. \eqref{BPS1}, \eqref{BPS2} and \eqref{BPS3} imply the
second order system in Eqs. \eqref{secondorder}. Moreover, one can check
that the BPS equations also imply the field equations obtained directly
varying the action in Eq. \eqref{I} with respect to the $U$ field and the
spatial components of the Maxwell potential $\vec{A}$.
\end{enumerate}

\subsection{Inclusion of the pions mass}

The present formalism can be applied even when the gauged $\chi $PT includes
a pions mass term. As it has been discussed many times in the literature, it
is possible to introduce a mass term for the pions in many different ways;
see, for instance, Refs. \cite{mass1}-\cite{mass5}. The reason is that, if one
write the chiral field as 
\begin{equation*}
U=\sigma \boldsymbol{1}+\pi _{i}t^{i}\ ,\qquad \sigma
^{2}+\sum_{i=1}^{3}\left( \pi _{i}\right) ^{2}=1\ ,
\end{equation*}%
the mass term for the pions in the lagrangian can be seen as an interaction
potential for $\sigma $, namely, $\mathcal{L}_{\text{mass}}=V\left( \sigma
\right) $, and the mass of the pions $m_{\pi }$ reads%
\begin{equation*}
m_{\pi }^{2}=-\left. \frac{\partial ^{2}V}{\partial \sigma ^{2}}\right\vert
_{\sigma =1}\ .
\end{equation*}%
It is worth emphasizing that it is impossible to distinguish between the
potentials by just knowing the pion mass (as it has been discussed in
details in the above references). Here we will consider a mass term proposed
in Refs. \cite{mass1} and \cite{mass2}, 
\begin{equation}
\mathcal{L}_{\text{mass}}=-\frac{2Km_{\pi }^{2}}{e^{2}}\biggl(1-\frac{1}{4}%
\text{Tr}[U]^{2}\biggl)\ ,  \label{mass}
\end{equation}%
which has several benefits. Such a mass term, at the same time, supports
domain walls solutions (hadronic layers) and (for small pions field) gives
the usual pions mass term. The appearance of hadronic layers at finite
density is confirmed both by numerical simulations and by astronomical
observations (see \cite{Dorso}-\cite{Schmitt1}, and references therein). For
the above reasons, there are very strong experimental evidence in favor of
the mass term in Eq. (\ref{mass}).

Using the parametrization in Eqs. \eqref{U} and \eqref{U1}, the mass term in
Eq. \eqref{mass} reads 
\begin{equation}
\mathcal{L}_{\text{mass}}=\frac{2Km_{\pi }^{2}}{e^{2}}\sin ^{2}(\alpha )\ ,
\label{mass2}
\end{equation}%
and the free energy density in Eq. \eqref{F1} becomes 
\begin{align}
e^{2}\digamma =& \frac{K}{2}\biggl\{\left( \overrightarrow{\nabla }\alpha
\right) ^{2}+\sin ^{2}(\alpha )\left( \overrightarrow{D}\Phi \right) ^{2}%
\biggl\}+\frac{1}{2}(\vec{B}^{2})+2K\mu _{I}^{2}\cos ^{2}(\alpha )+2Km_{\pi
}^{2}\sin ^{2}(\alpha )-2K\mu _{I}^{2}  \notag \\
=& \frac{K}{2}\biggl\{\left( \overrightarrow{\nabla }\alpha \right)
^{2}+\sin ^{2}(\alpha )\left( \overrightarrow{D}\Phi \right) ^{2}\biggl\}+%
\frac{1}{2}(\vec{B}^{2})+2K\left( \mu _{I}^{2}-m_{\pi }^{2}\right) \cos
^{2}(\alpha )-2K\left( \mu _{I}^{2}-m_{\pi }^{2}\right) \ .  \label{FMASS}
\end{align}%
Consequently, the inclusion of the mass term for the pions manifests itself
in a negative shift\footnote{%
It is worth noting that the combination $\mu _{I}^{2}-m_{\pi }^{2}$ is quite natural in this context (see for
instance \cite{IsospinR7} and references therein).} of the $%
\mu _{I}^{2}$:%
\begin{equation*}
\mu _{I}^{2}\rightarrow \mu _{I}^{2}-m_{\pi }^{2}\ .
\end{equation*}%
Hence, all the results in the previous sections still hold provided the
shift in the above equation is taken into account. The critical value for
the isospin chemical potential when the pions mass is taken into account
reads 
\begin{equation} \label{boundmu}
\mu _{I}^{\text{c}}=\sqrt{K+m_{\pi }^{2}}<1.1m_{\pi }\ .
\end{equation}%
It is important to emphasize that, even with the inclusion of the mass term,
our value for the critical isospin chemical potential $\mu _{I}^{\text{c}}$
are within the range of validity of $\chi $PT (namely, $0.1\,m_{\pi
}\lesssim \mu _{I}^{\text{c}},<1.1m_{\pi }$). In fact, it is known (see \cite%
{IsospinR7}, and references therein) that $\chi $PT needs improvement
starting from $\mu _{I}\sim 1.3m_{\pi }$.

%%%%%%%%%%%%%%%%%%%%%%%%%%%%%%%%%%%%%%%%%%%%
\section{A single vortex} \label{sec-5}
%%%%%%%%%%%%%%%%%%%%%%%%%%%%%%%%%%%%%%%%%%%%

In this section, we will study the case of a single vortex. This is great
interest because it allows to explicitly show many of the most important
properties of our solution.

The starting point is the pionic field in Eq. \eqref{singlevortex1} together
with the metric in Eq. \eqref{singlevortex2}. The Gibbs free energy density
for a single vortex reduces to 
\begin{align*}
e^2 G=& \frac{K}{2r^{2}}\{r^{2}(\alpha')^2+\sin ^{2}(\alpha
)(n-2A)^{2}\}+\frac{1}{2r^{2}}(A')^2+2K\mu _{I}^{2}\cos ^{2}(\alpha) \\
=& \frac{K}{2r^{2}}\{r\alpha'+\sin(\alpha)(n-2A)\}^2  +\frac{1}{2r^{2}}\{A'+\bar{\mu} r\cos (\alpha )\}^{2}-K\frac{1}{r}\sin (\alpha )(n-2A)\alpha'-\bar{\mu}\frac{1}{r}\cos (\alpha )A'\ ,
\end{align*}
where we have defined $\bar{\mu}^{2}=4K\mu _{I}^{2}$. Now, we can follow the
same steps presented in the previous sections in a straightforward way.
Integrating the last expression, we obtain the Gibbs free energy 
\begin{align} \notag
\frac{\boldsymbol{G}}{L}=& 2\pi \int \frac{1}{2e^2r}\{K(r\alpha'+\sin(\alpha)(n-2A))^2+ (A'+\bar{\mu} r\cos (\alpha ))^2   \}dr \\
& -\frac{2\pi}{e^2}\int \left\{ K\sin (\alpha )(n-2A)\alpha'+\bar{\mu}%
\cos (\alpha )A'\right\} dr \ . \label{G2}
\end{align}
The second term in Eq. \eqref{G2} is a total derivative for the particular
value of the isospin chemical potential presented in the previous section,
that is $\bar{\mu}^{\text{c}}=2K\Rightarrow \mu _{I}^{\text{c}}=\sqrt{K}$.
In fact, for this value
\begin{equation*}
-K\sin (\alpha )(n-2A)\alpha ^{\prime }-\bar{\mu}\cos (\alpha )A^{\prime
}=\{K\cos (\alpha )(n-2A)\}^{\prime }\ .
\end{equation*}%
It follows from Eq. \eqref{G2} that the Gibbs free energy is minimized when
the following BPS equations are satisfied 
\begin{gather}
r\alpha ^{\prime }+\sin (\alpha )(n-2A)=0\ ,  \label{e1} \\
A^{\prime }+2Kr\cos (\alpha )=0\ ,  \label{e2}
\end{gather}%
where, also in this case, one can check that the above first order BPS
system implies the second-order field equations corresponding to the
stationary points of the Gibbs free energy in Eq. \eqref{G2}. 

The topological charge of this configuration reads
\begin{equation}
\omega=  \frac{2 \pi}{e^2} \int \{ K\cos(\alpha)(n-2A)  \}' dr =  \frac{2 \pi K }{e^2} n \ . \label{Q}
\end{equation}
Note that, as we have pointed out in the previous section, the topological charge density of the vortices here presented is different from the magnetic flux density (due to the presence of the term $\cos(\alpha)$ in Eq. \eqref{Q}). However, once the boundary conditions are implemented, the topological charge is in fact proportional to the magnetic flux (times a dimensional constant that depends on the couplings of the theory), as can be seen from Eqs. \eqref{fluxquantization} and \eqref{Q}. Now, if the solitons are confined in a finite volume, the topological charge and the magnetic flux may be different.\footnote{For instance, if one analyzes this system on a finite sample, one does not necessarily need to require that $\cos(\alpha)=0$ on the boundary.} This more general case will be studied in a forthcoming paper. 

\begin{figure}[tbp]
         \centering \includegraphics[width=0.6\textwidth]{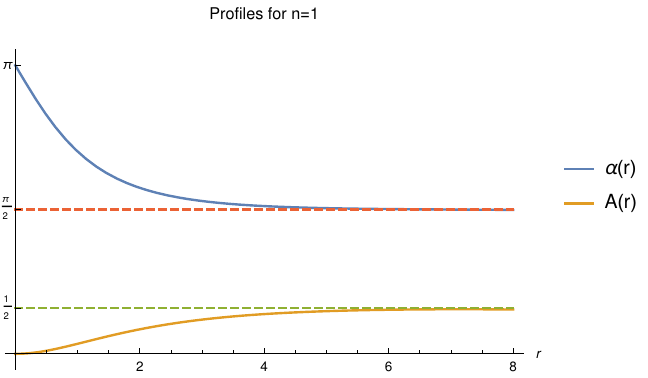}
\caption{Profiles $\alpha$ and $A$ of the single vortex with $n=1$.}
\label{Fig-profiles}
\end{figure}

\begin{figure}[hbtp]
\centering
\includegraphics[width=0.45\textwidth]{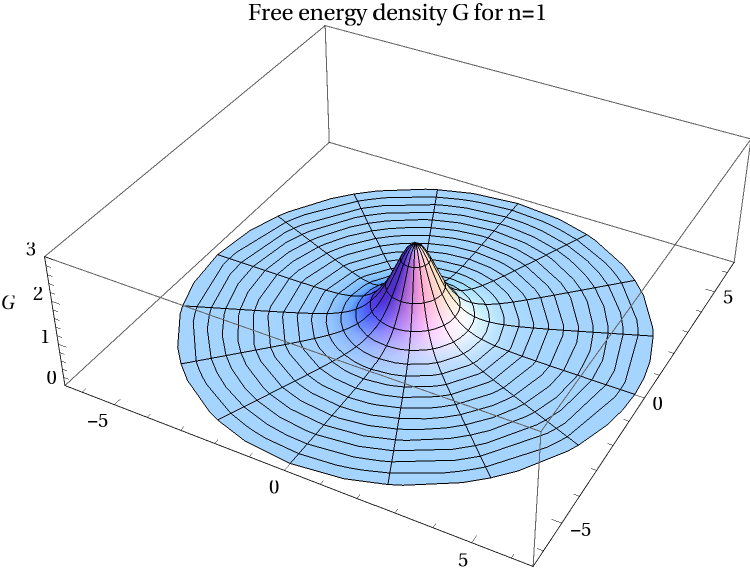}
\quad 
\includegraphics[width=0.45\textwidth]{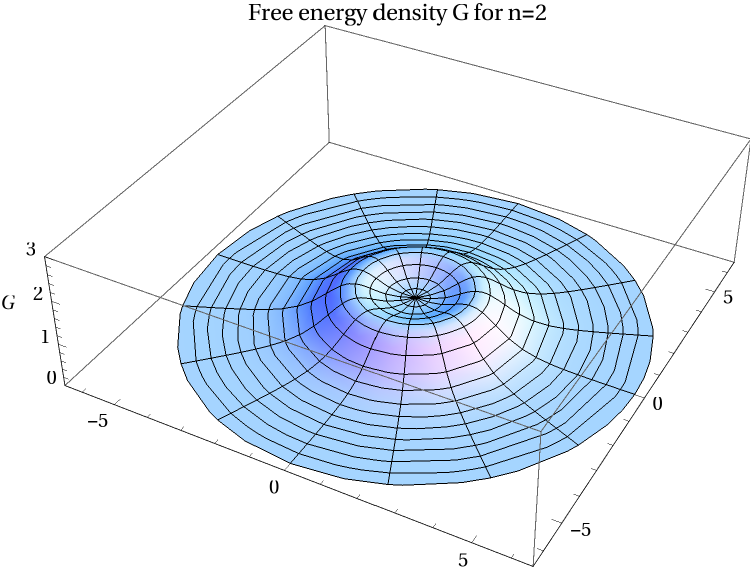}
\quad 
\includegraphics[width=0.45\textwidth]{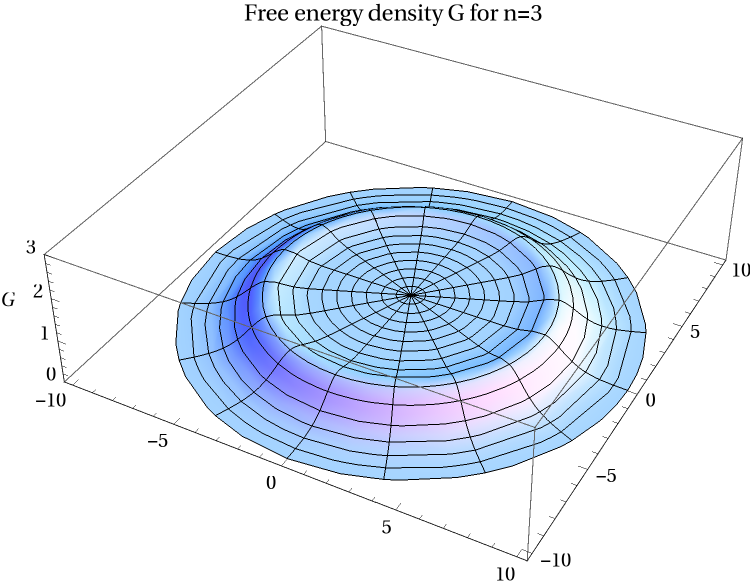}
\caption{Gibbs free energy density for a single vortex with $n=1$, $n=2$ and $n=3$.}
\label{Fig-G}
\end{figure}

Fig. \ref{Fig-profiles} shows the behavior of the pionic profile $\alpha$ as well as the Maxwell potential characterized by the function $A$ for the single vortex with $n=1$. 
To integrate the system in Eqs. \eqref{e1} and \eqref{e2} we have used the boundary conditions in Eqs. \eqref{bc1} and \eqref{bc2}, with $c_1=0$ and $c_2=1$. To obtain clearer plots, we have fixed the value of the coupling constants to $K=\frac{1}{7}$ and $f_\pi=2$. 
Fig. \ref{Fig-G} shows the Gibbs free energy density of a single vortex for different values of $n$. 

\begin{figure}[tbp]
\centering \includegraphics[width=0.6\textwidth]{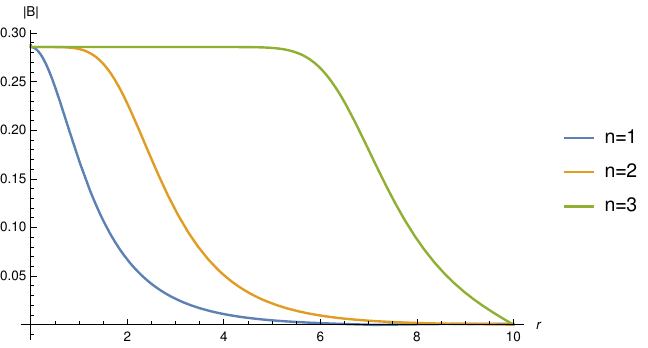}
\caption{Magnetic field of the single vortex for for $n=1$, $n=2$ and $n=3$.}
\label{Fig-B}
\end{figure}

Regarding the electromagnetic field self-sustained by the vortex, the
magnetic field is in the $z$ direction, given by 
\begin{equation}
B_{z}=-\frac{A^{\prime }}{r} \ ,   \label{Bz}
\end{equation}%
while the current in the $\theta $ direction is 
\begin{equation}
J_{\theta }=2K\sin ^{2}(\alpha )(n-2A)\ .
\end{equation}%
Fig. \ref{Fig-B} shows the magnitude of the magnetic field of the single vortex for different values of $n$.

%%%%%%%%%%%%%%%%%%%%%%%%%%%%%%%%%%%%%%%%%%%%
\section{Final remarks} \label{sec-6}
%%%%%%%%%%%%%%%%%%%%%%%%%%%%%%%%%%%%%%%%%%%%

In the present manuscript, we have shown that, at a critical value of the
isospin chemical potential, superconducting multi-vortices configurations
appear in gauged $\chi$PT. This has been achieved through the derivation of a novel BPS
bound in $\chi $PT minimally coupled to electrodynamics, which is saturated
by these multi-vortices configurations. This bound has not been discovered
before because the topological charge density is not the magnetic flux
density but rather is a magnetic flux density \say{dressed} by the hadronic
profile. Such a dressing discloses the screening effects of the hadrons on
the magnetic field, allowing us to derive an upper bound on the intensity of the
magnetic field itself. These configurations are supported by a persistent
current, which does not vanishes when the Maxwell field is suppressed,
showing that the condensate has superconducting characteristics. These BPS
vortices exist for a particular value of the isospin chemical potential
which is within the range of validity of $\chi $PT. The present results open
many interesting questions on which we hope to come back in a future
publication. First of all, in the pioneering paper \cite{Weinberg}, E.
Weinberg, using in a very clever way the index theorem, was able to
determine the dimension of the moduli space of BPS multi-vortices in the GL
theory at critical coupling. Then, there are two natural questions. The first is
whether the dimension in the present case is the same as in the GL theory
(the arguments presented in the Appendix suggest a positive answer). The
second is whether the index appearing in the present case is related to the
magnetic flux density, or the dressed magnetic flux density. Another closely
related direction of research is the remarkable idea proposed by N. Manton
(see \cite{Manton1}-\cite{Samols}, and
references therein), according to which one can describe the dynamics of BPS
solitons in terms of the geodesics of the geometry of the moduli space. Hence,
it is an extremely interesting issue to analyze the geometry of the moduli
space in the present case. On a more phenomenological side, the analysis
of the coupling of up and down quarks to these BPS multi-vortices is very relevant.
We think that all the above open questions deserve further investigation.

\acknowledgments

F. C. has been funded by FONDECYT Grant No. 1240048. M. L. has been funded
by FONDECYT Iniciaci\'on No. 11241079.

\appendix

\section{On the existence of multi-vortices configurations}

In this Appendix, we outline two mathematical strategies (which can be
promoted to rigorous arguments, as it will be discussed in a future
publication) leading to the proof of the existence of multi-vortices
configurations in the $\chi $PT coupled to Maxwell at the critical value of
the isospin chemical potential defined in the previous sections. Besides the
intrinsic importance to prove rigorously the existence of multi-vortices
solutions, the first strategy presented in this Appendix shows in a very
clear way the difference between the present case, in which the hadronic
profile screens substantially the magnetic field, and the usual GL case.

\subsection{First Strategy}

As it has been explained in the previous sections, the BPS equations in Eqs.
(\ref{BPS1}), (\ref{BPS2}) and (\ref{BPS3}) are equivalent to the master
equation in Eq. \eqref{MasterEquation} which is a semi-linear elliptic
equation in two coordinates. Moreover, in the case of multi-vortices in
the GL theory at critical coupling, the BPS equations are equivalent to a
single master equation, which is also a semi-linear elliptic equation in two
coordinates (see \cite{Taubes1}, \cite{Taubes2}, and references therein).
The differences between the two master equations only lie in the non-linear
terms. Thus, let us write the two master equations in a unified fashion: 
\begin{gather}
\nabla ^{2}H=\Psi _{I}\left( H\right) +\sum_{j=1}^{N}n_{j}\delta \left(
x-x_{j},y-y_{j}\right) \ ,\ \ I=1,2\ ,  \label{mastere1} \\
\Psi _{1}\left( H\right) =K\tanh (H)\ ,\qquad \Psi _{2}\left( H\right)
=-1+\exp H\ ,  \notag
\end{gather}%
where $\Psi _{1}\left( H\right) $ corresponds to the BPS vortices in $\chi $%
PT presented in this paper while $\Psi _{2}\left( H\right) $ corresponds to
the BPS vortices in GL theory at critical coupling. Now, we will describe a
technique (which is quite standard in the theory of non-linear elliptic
equations; see Ref. \cite{GilbargTrudinger}) based on the Banach-Caccioppoli
theorem (see Ref. \cite{BanachCaccioppoli1} and references therein). The
idea to use a strategy based on the Banach-Caccioppoli theorem is that it
highlights a very important difference between $\Psi _{1}\left( H\right) $
and $\Psi _{2}\left( H\right) $ which, physically, is related to the
screening of the magnetic field from the hadronic profile, which is stronger
in $\chi $PT than in GL theory at critical coupling. The mathematical
consequence is that in the case of $\Psi _{1}\left( H\right) $ one can apply
directly the Banach-Caccioppoli theorem, while in the case of $\Psi
_{2}\left( H\right) $ the existence and uniqueness parts require far more
refined techniques (see \cite{Taubes1}, \cite{Taubes2}, and references
therein).

Let us begin with the definition of an auxiliary linear problem: 
\begin{gather}
\nabla ^{2}u = \Psi _{I}\left( V\right) +\sum_{j=1}^{N}n_{j}P_{j}\ ,\
n_{j}\in \mathbb{N} \ ,\ \ \left. u\right\vert _{\partial \Omega }=0\ ,\ \
V\in X\ ,\ N<\infty \ ,\   \label{aux1} \\
\Psi _{1}\left( V\right) = K\tanh V\ ,\quad \Psi _{2}\left( V\right)
=-1+\exp V\ ,\quad \nabla ^{2}=\frac{\partial ^{2}}{\partial x^{2}}+\frac{%
\partial ^{2}}{\partial y^{2}}\ ,  \label{aux1.1}
\end{gather}%
where the functions $P_{j}$ are defined as follows:%
\begin{gather}
P_{j}\left( x,y\right) =\Lambda \ ,\ \ \left( x,y\right) \in \Omega _{j},\ \
P_{j}\left( x,y\right) =0\ ,\ \ \left( x,y\right) \in \Omega \diagdown
\Omega _{j}\ ,  \label{aux1.2} \\
\text{measure} \left( \Omega _{j}\right) =\frac{1}{\Lambda ^{2}}\ ,\ \
\Omega _{j}\subset \Omega \ \forall j=1,..,N\ ,\ \ \Omega _{j}\cap \Omega
_{k}=0\ for\ j\neq k\ ,  \label{aux1.3} \\
P_{j}\ \in L^{2}\left( \Omega \right) \ \ \ \forall j=1,..,N\ ,
\label{aux1.4} \\
\Omega =\left( 0,a\right) \times \left( 0,a\right) \ .  \label{aux1.5}
\end{gather}%
The conditions in Eqs. (\ref{aux1.2}), (\ref{aux1.3}) and (\ref{aux1.4})
describe the functions $P_{j}$ (where $\Omega _{j}$ is a small circle of
area $\frac{1}{\Lambda ^{2}}$ around the point of coordinates $(x_{j},y_{j})$%
). It is clear that when $\Lambda $ is a large number, the $P_{j}$ tend to
be very peaked $(x_{j},y_{j})$: the idea to introduce the functions $P_{j}$
is to define a limit in which, as $\Lambda $ becomes bigger and bigger, $%
P_{j}$ becomes closer and closer to a Dirac delta. There are actually many
different ways to represent the Dirac delta as the limit of bounded (even
smooth) functions. This issue will be discussed in more detail in a future
publication.

In the linear elliptic equation in Eq. (\ref{aux1}), $u=u(x,y)$ is the
unknown, while $V=V(x,y)$ is supposed to be known. For the moment, we will
take as domain $\Omega $ a square of side $a<K^{-1/2}$ in Eq. (\ref{aux1.5})
(then we will explain how to obtain similar results in bigger squares).

\textit{The first technical step} is to choose the function $V$ (the
\textquotedblleft source term" of the auxiliary problem) in a suitable
functional space $X$ in such a way that the solution $u(x,y)$ of the
auxiliary problem belongs to the same functional space of the function $V$.

\textit{The second technical step} is to define a mapping $\Phi $ such that
for any $V_{k}\in X$ the mapping $\Phi $ sends $V_{k}$ into the solution $%
u_{k}$ of the auxiliary problem in Eq. (\ref{aux1}) with $V=V_{k}$: 
\begin{equation*}
u_{k}=\Phi \left( V_{k}\right) |\qquad \nabla ^{2}u_{k}=\Psi _{I}\left(
V_{k}\right) +\sum_{j=1}^{N}n_{j}P_{j}\ ,\ n_{j}\in \mathbb{N}\ ,\ \ \left.
u_{k}\right\vert _{\partial \Omega }=0\ .
\end{equation*}

\textit{The third technical step} is to show that such a mapping is a
contraction so that, due to the Banach-Caccioppoli theorem, it possesses a
unique fixed point. Then, it is trivial to see that such a fixed point is a
solution of the original non-linear problem in Eq. (\ref{mastere1}).

\subsubsection{First step}

Here we need to put the basis to define such a mapping. The easiest choice
is to assume that the functional space $X$ (to which the source $V$ in Eq. (%
\ref{aux1}) belongs) is $L^{2}\left( \Omega \right) $:%
\begin{equation*}
X=L^{2}\left( \Omega \right) \ ,
\end{equation*}%
where $L^{2}\left( \Omega \right) $ is the space of square integrable
function on $\Omega $. Moreover, one observes the following. If%
\begin{equation}
\nabla ^{2}u=f\ ,\ \ \left. u\right\vert _{\partial \Omega }=0\ ,\ \ f\in
L^{2}\left( \Omega \right) \Rightarrow \ u\in L^{2}\left( \Omega \right) \ ,
\label{aux2.1}
\end{equation}%
then (due to $\left. u\right\vert _{\partial \Omega }=0$) one gets%
\begin{gather}
\int_{\Omega }uf=\int_{\Omega }u\nabla ^{2}u=-\left[ \int_{\Omega }\left( 
\frac{\partial u}{\partial x}\right) ^{2}+\int_{\Omega }\left( \frac{%
\partial u}{\partial y}\right) ^{2}\right]  \notag \\
\Rightarrow \left[ \int_{\Omega }\left( \frac{\partial u}{\partial x}\right)
^{2}+\int_{\Omega }\left( \frac{\partial u}{\partial y}\right) ^{2}\right]
\leq \left\Vert f\right\Vert _{L^{2}\left( \Omega \right) }\left\Vert
u\right\Vert _{L^{2}\left( \Omega \right) }\ ,  \label{A2}
\end{gather}%
where we have used the Schwarz inequality (which will be used often in the
following computations). Using the obvious identity%
\begin{equation*}
\int_{0}^{x}\frac{\partial u}{\partial \sigma }\left( \sigma ,y\right)
d\sigma =u(x,y)\ ,
\end{equation*}%
it follows that for any $u(x,y)$ which solves the problem in Eq. (\ref%
{aux2.1}) one gets%
\begin{gather}
\left( \left\Vert u\right\Vert _{L^{2}\left( \Omega \right) }\right)
^{2}=\int_{\Omega }u^{2}dxdy=\int_{\Omega }\left[ \int_{0}^{x}\frac{\partial
u}{\partial \sigma }\left( \sigma ,y\right) d\sigma \right] ^{2}dxdy  \notag
\\
\Rightarrow \left( \left\Vert u\right\Vert _{L^{2}\left( \Omega \right)
}\right) ^{2}\leq \int_{\Omega }\left[ \left( \int_{0}^{x}\left( \frac{%
\partial u}{\partial \sigma }\left( \sigma ,y\right) \right) ^{2}d\sigma
\right) ^{1/2}\left( \int_{0}^{x}1^{2}d\sigma \right) ^{1/2}\right] ^{2}dxdy%
\overset{def}{=}\Upsilon \ .  \label{aux2.2}
\end{gather}%
It is possible to further bound from above $\Upsilon $\ as follows:%
\begin{eqnarray*}
\Upsilon &\leq &a\int_{\Omega }\left( \int_{0}^{a}\left( \frac{\partial u}{%
\partial \sigma }\left( \sigma ,y\right) \right) ^{2}d\sigma \right)
dxdy=a\int_{0}^{a}dx\left[ \int_{0}^{a}dy\left( \int_{0}^{a}\left( \frac{%
\partial u}{\partial \sigma }\left( \sigma ,y\right) \right) ^{2}d\sigma
\right) \right] \\
&=&a\int_{0}^{a}dx\left[ \int_{\Omega }\left( \frac{\partial u}{\partial x}%
\right) ^{2}dxdy\right] dx=a^{2}\int_{\Omega }\left( \frac{\partial u}{%
\partial x}\right) ^{2}dxdy\ ,
\end{eqnarray*}%
\begin{equation}
\Rightarrow \left( \left\Vert u\right\Vert _{L^{2}\left( \Omega \right)
}\right) ^{2}\leq a^{2}\int_{\Omega }\left( \frac{\partial u}{\partial x}%
\right) ^{2}dxdy\ .\   \label{A3}
\end{equation}%
Hence, from Eqs. \eqref{A2} and \eqref{A3} it follows that%
\begin{gather}
\int_{\Omega }\left( \frac{\partial u}{\partial x}\right) ^{2}\leq a\left(
\int_{\Omega }\left( \frac{\partial u}{\partial x}\right) ^{2}dxdy\right)
^{1/2}\left\Vert f\right\Vert _{L^{2}\left( \Omega \right) }  \notag \\
\Rightarrow \left( \int_{\Omega }\left( \frac{\partial u}{\partial x}\right)
^{2}dxdy\right) ^{1/2}\leq a\left\Vert f\right\Vert _{L^{2}\left( \Omega
\right) }\ ,  \label{A4}
\end{gather}%
and this is true $\forall \ u$ which is a solution of the problem in Eq. (%
\ref{aux2.1}). Similarly, we get%
\begin{equation}
\left( \int_{\Omega }\left( \frac{\partial u}{\partial y}\right)
^{2}dxdy\right) ^{1/2}\leq a\left\Vert f\right\Vert _{L^{2}\left( \Omega
\right) }\ .  \label{A5}
\end{equation}%
Moreover, from Eqs. (\ref{A3}) and (\ref{A4}) together with (\ref{A5}), we
also get%
\begin{equation}
\left\Vert u\right\Vert _{L^{2}\left( \Omega \right) }\leq a^{2}\left\Vert
f\right\Vert _{L^{2}\left( \Omega \right) }\ .  \label{A4.1}
\end{equation}%
As it is well known (see, for instance, \cite{GilbargTrudinger}) the problem
in Eq. (\ref{aux2.1}) possesses an existence and uniqueness theorem: $%
\forall f\in L^{2}\left( \Omega \right) $ there is a unique $u$ which solves
the problem in Eq. (\ref{aux2.1}). One can also say that, actually, the
solution $u$ actually belongs to a well-behaved subspace of $L^{2}\left(
\Omega \right) $:%
\begin{equation*}
u\in \mathbb{W\ },\qquad \mathbb{W}\overset{def}{=}\left\{ u\in
W^{2,2}\left( \Omega \right) ,\ \left. u\right\vert _{\partial \Omega
}=0\right\} \ ,
\end{equation*}%
where $W^{2,2}\left( \Omega \right) $\ is the (2,2) Sobolev space on the
domain $\Omega $ (for a definition of Sobolev space, see \cite%
{GilbargTrudinger} and references therein).

\subsubsection{Second step}

Here we define such a mapping. As a particular case of the problem in Eq. (%
\ref{aux2.1}), let us consider the problem in Eq. (\ref{aux1}) with $\Psi
_{1}\left( V\right) =K\tanh V$, where%
\begin{equation*}
V\in L^{2}\left( \Omega \right) \ .
\end{equation*}%
Thus, in this case, 
\begin{equation*}
f=K\tanh V+\sum_{j=1}^{N}n_{j}P_{j}\in L^{2}\left( \Omega \right) \ .
\end{equation*}%
The existence and uniqueness result for the problem in Eq. (\ref{aux2.1})
provides with the following:%
\begin{equation*}
\forall \, V\in L^{2}\left( \Omega \right) \ ,\ \left. \exists \right\vert \
u\in \mathbb{W\subset \ }L^{2}\left( \Omega \right) \ ,
\end{equation*}%
which solves the problem in Eq. (\ref{aux1}) with $\Psi _{1}\left( V\right)
=K\tanh V$. We are in a position to define a mapping $\Phi $ such that for
any $V\in L^{2}\left( \Omega \right) $ the mapping $\Phi $ sends $V$ into
the unique solution $u$ of the auxiliary problem in Eq. (\ref{aux1}): 
\begin{gather}
u=\Phi \left( V\right) |  \label{mapping1} \\
\nabla ^{2}u=\Psi _{1}\left( V\right) +\sum_{j=1}^{N}n_{j}P_{j}\ ,\ n_{j}\in 
\mathbb{N}\ ,\quad u|_{\partial \Omega }=0\ ,\quad \Psi _{1}\left( V\right)
=K\tanh V\ .  \label{mapping2}
\end{gather}%
Let us consider now two different \textquotedblleft sources" $V_{1}$ and $%
V_{2}$ together with the corresponding images under the mappings $u_{1}=\Phi
\left( V_{1}\right) $ and $u_{2}=\Phi \left( V_{2}\right) $. From the above
definitions it follows that%
\begin{equation*}
\nabla ^{2}\left( u_{1}-u_{2}\right) =K\left( \tanh V_{1}-\tanh V_{2}\right)
\in L^{2}\left( \Omega \right) \ ,\ \ \left. \left( u_{1}-u_{2}\right)
\right\vert _{\partial \Omega }=0\ ,\quad u_{1}-u_{2}\in \mathbb{W\ }.
\end{equation*}%
Then, from Eq. (\ref{A4.1}), it follows that%
\begin{gather}
\left\Vert u_{1}-u_{2}\right\Vert _{L^{2}\left( \Omega \right) }\leq
a^{2}\left\Vert K\left( \tanh V_{1}-\tanh V_{2}\right) \right\Vert
_{L^{2}\left( \Omega \right) }  \notag \\
\Rightarrow \left\Vert u_{1}-u_{2}\right\Vert _{L^{2}\left( \Omega \right)
}\leq Ka^{2}\left\Vert \tanh V_{1}-\tanh V_{2}\right\Vert _{L^{2}\left(
\Omega \right) }\ =Ka^{2}\left( \int_{\Omega }\left\vert \tanh V_{1}-\tanh
V_{2}\right\vert ^{2}dxdy\right) ^{1/2}\ .  \label{mapping2.1}
\end{gather}%
Then, using the elementary result from calculus (namely, the mean value
theorem) that for any continuous differentiable function $h(x)$ we have%
\begin{eqnarray}
h(x)-h(y) &=&h^{\prime }(z)\left( x-y\right) \ ,\ h^{\prime }(z)=\frac{dh}{dz%
}\ ,\ \ z\in \left( x,y\right) \   \notag \\
\Rightarrow \left\vert \tanh V_{1}-\tanh V_{2}\right\vert &=&\frac{1}{\cosh
^{2}\left( z\right) }\ \left\vert V_{1}-V_{2}\right\vert \ ,\ \ \frac{1}{%
\cosh ^{2}\left( z\right) }\leq 1\ \forall z\in 
\mathbb{R}
\ ,  \label{mean}
\end{eqnarray}%
for some $z$. Consequently, from Eq. (\ref{mapping2.1}) together with the
mean value theorem (and taking into account that the derivative of the
hyperbolic tangent is always less than or equal to 1 in absolute value) we
get%
\begin{equation}
\left\Vert u_{1}-u_{2}\right\Vert _{L^{2}\left( \Omega \right) }\leq
Ka^{2}\left( \int_{\Omega }\left\vert V_{1}-V_{2}\right\vert ^{2}dxdy\right)
^{1/2}\ .  \label{mapping3}
\end{equation}

\subsubsection{Third step}

Here we prove that the mapping defined in Eqs. (\ref{mapping1}) and (\ref%
{mapping2}) is a contraction thanks to Eq. (\ref{mapping3}). Indeed, from
the definition in Eqs. (\ref{mapping1}) and (\ref{mapping2}) together with
Eq. (\ref{mapping3}) we obtain that%
\begin{eqnarray}
\forall \ V_{1}\ ,\ V_{2}\ &\in &L^{2}\left( \Omega \right) \ ,  \notag \\
\left\Vert \Phi \left( V_{1}\right) -\Phi \left( V_{2}\right) \right\Vert
_{L^{2}\left( \Omega \right) } &\leq &Ka^{2}\left\Vert
V_{1}-V_{2}\right\Vert _{L^{2}\left( \Omega \right) }\ .  \label{mapping4}
\end{eqnarray}%
Consequently, if we take $Ka^{2}<1$, we see that the mapping defined here
above is a contraction so that, due to the Banach-Caccioppoli theorem, it
possesses a unique fixed point:%
\begin{equation}
\left. \exists \right\vert V^{\ast }\quad \text{such that}\quad \Phi \left(
V^{\ast }\right) =V^{\ast }\ .  \label{fixed point}
\end{equation}%
If we explicitly write the condition to be a fixed point of the mapping
defined in Eqs. (\ref{mapping1}) and (\ref{mapping2}), we see that $V^{\ast
} $ satisfies the following equation:%
\begin{equation*}
\nabla ^{2}V^{\ast }=\Psi _{1}\left( V^{\ast }\right)
+\sum_{j=1}^{N}n_{j}P_{j}\ ,\ n_{j}\in 
\mathbb{N}
\ ,\ \ \left. V^{\ast }\right\vert _{\partial \Omega }=0\ ,\ \ \Psi
_{1}\left( V^{\ast }\right) =K\tanh V^{\ast }\ ,
\end{equation*}%
namely, the fixed point $V^{\ast }$ of the mapping defined here above is the
solution of the sought non-linear elliptic differential equation. Note that
such simple argument would not work in the case of the BPS vortices in the
GL theory. The reason is that, in the GL case, $\Psi _{2}\left( V\right)
=-1+\exp V$, and if we follow the same procedure, when we apply the mean
value theorem we would get%
\begin{equation*}
\left\vert \exp V_{1}-\exp V_{2}\right\vert =\left( \exp z\right) \
\left\vert V_{1}-V_{2}\right\vert \ ,
\end{equation*}%
for some $z$. However, unlike what happens with the $\Psi _{1}\left(
V\right) =K\tanh V$, which characterizes the BPS vortices in $\chi $PT
introduced in the present work, $\left\vert \exp z\right\vert $ is not less
than or equal to one. This is the reason why the proof of existence and
uniqueness of BPS multi-vortices in the GL case requires very refined
techniques (see Refs. \cite{Taubes1} and \cite{Taubes2}). This technical
difference is related to the fact that, in the case of $\chi $PT, the
hadronic profile screens considerably the magnetic field.

It is worth emphasizing that the above argument is not complete. First of
all, one should define properly the limit of $\Lambda \rightarrow \infty $
in order to obtain $P_{j}\rightarrow \delta (x-x_{j},y-y_{j})$. This issue
appears to be within the reach of usual techniques in functional analysis
(see Refs. \cite{GilbargTrudinger} and \cite{BanachCaccioppoli1}). Secondly,
the smoothness of the solution must be improved, as we would like $V^{\ast }$
to be smooth, far from the positions of the vortices. The usual \say{bootstrap
arguments} or \say{elliptic regularity theory} in Sobolev spaces (which have
been already used in the GL case: see \cite{YsonnYang}-\cite{Taubes2}) appear to work in the present case as well without much
change. Thirdly, we would like to eliminate the constraint $Ka^{2}<1$ on the
size of the domain $\Omega $ allowing domains of generic sizes. This can be
certainly be achieved using more refined inequalities in functional analysis
(see \cite{BanachCaccioppoli2}\ and references therein).

It is important to remark here that the idea of the present paper is not to
give a fully rigorous proof of the existence and uniqueness (for this a
separate more mathematical paper is necessary) but to provide with sound
evidences that such BPS multi-vortices configurations do exist. In the
present work we have discussed (in this Appendix) not only strong
mathematical arguments in favor of the existence of such BPS configurations
but also (in the main text) we have constructed numerical solutions for the
single vortex case.

\subsection{Second Strategy}

The first strategy (which has been presented here above) is interesting
since it emphasizes how the physical properties, which has been discussed in
the main text (such as the hadronic screening of the magnetic field)
manifest themselves in the possibility of using the contraction theorem
directly. However, the easiest and fastest strategy to prove rigorously that
such multi-vortices exist is actually to follow exactly the same steps of Ref. 
\cite{Taubes1}. A direct analysis reveals that many of the steps can be
repeated without change in the present case. In particular, we can start
from our main equation 
\begin{equation}
\nabla ^{2}H=K\tanh (H)+\sum_{j=1}^{N}n_{j}\delta \left(
x-x_{j},y-y_{j}\right) \ .  \label{masterlast}
\end{equation}%
Then, the functions $u_{0}$ and $g_{0}$ in Eqs. (3.2) and (3.4) of Ref. \cite%
{Taubes1} can be introduced in the present case as well in order to take
care of the Dirac deltas. More precisely, one can choose (the constant
parameters appearing in) $u_{0}$ in such a way that its Laplacian cancels
exactly the sum over the Dirac deltas, leaving the smooth function $g_{0}$
only. Then, (as in Eq. (3.5) of Ref. \cite{Taubes1}) the new unknown $v$ can be
defined as follows:%
\begin{equation*}
H=u_{0}+v\ .
\end{equation*}%
If the above expression is replaced into Eq. (\ref{masterlast}) one arrives
at%
\begin{equation*}
\nabla ^{2}v=g_{0}+K\tanh (u_{0}+v)\ .
\end{equation*}%
The above equation for $v$ can be derived by the following action%
\begin{equation}
\mathcal{S}_{\text{eff}}(v)=\frac{1}{2}(\vec{\nabla}v)^{2}+g_{0}v+K\ln {%
(2\cosh (u_{0}+v))}\ ,  \label{masteraction}
\end{equation}%
to be compared with Eq. (3.7) of Ref. \cite{Taubes1}. The most important step in the analysis of
Ref. \cite{Taubes1} is the proof that the action in his Eq. (3.7) is a
convex functional. In particular, in Eq. (4.28) of \cite{Taubes1} a
fundamental role is played by the strict convexity of the non-linear part of
the action (the exponential in that case). On the other hand, one can
observe that the term $K\ln {(2\cosh (u_{0}+v))}$, in Eq. (\ref{masteraction}%
), is a strict convex function of its argument as well. Moreover, there are
many technical steps of Ref. \cite{Taubes1} which actually do not depend on
the explicit form of the non-linear term in the action in his Eq. (3.7):
such steps can be imported directly to the present situations. At last,
there are steps (such as the inequalities in Eqs. (4.6) and (4.7) of \cite%
{Taubes1}) which can be easily generalized to the present case thanks both
to Eq. (\ref{mean}) discussed in the first part of the Appendix and to the
fact that $\ln {(2\cosh X)}$ is well approximated by $\left\vert
X\right\vert $. 

The above sketchy arguments show that the strategy of Ref. \cite{Taubes1}
can be adapted to the present case, leading to a rigorous proof of the
existence of multi-vortex solutions. We will come back to this issue in a
future publication.


\begin{thebibliography}{99}

\bibitem{R0} P. de Forcrand, \textit{Simulating QCD at finite density},
PoS(LAT2009)010 [arXiv:1005.0539] [INSPIRE].

\bibitem{R11} J. B. Kogut, M. A. Stephanov, The phases of Quantum
Chromodynamics, Cambridge University Press (2004).

\bibitem{R2} N. Brambilla et al., QCD and Strongly Coupled Gauge Theories:
Challenges and Perspectives, Eur. Phys. J. C 74 (2014) 2981
[arXiv:1404.3723] [INSPIRE].

\bibitem{Pisarski1} R.~Pisarski, 
%``Three Lectures on QCD Phase Transitions,''
Lect. Notes Phys. \textbf{999}, 89-145 (2022).

\bibitem{sign1} A. Bzdak, S. Esumi, V. Koch, J. Liao, M. Stephanov et al.,
Phys. Rept. 853 (2020) 1-87.

\bibitem{sign2} K. Nagata, Prog. Part. Nucl. Phys. 127 (2022) 103991.

\bibitem{sign3m} N. Astrakhantsev, V.V. Braguta, N.V. Kolomoyets, A.Yu.
Kotov, D.D. Kuznedelev et al., Phys.Part.Nucl. 52 (2021) 4, 536-541; N.
Astrakhantsev, V. Braguta, M. Cardinali, M. D.Elia, L. Maio et al., PoS
LATTICE2021 (2022) 119; B. B. Brandt, F. Cuteri, G. Endr\H{o}di, G. Mark\'o,
L. Sandbote, A. D. M. Valois, arXiv:2305.19029.

\bibitem{HIC} W. Busza, K. Rajagopal, W. van der Schee, Annu. Rev. Nucl.
Part. Sci. 2018. 68:339--76.

\bibitem{QGBook} K. Yagi, T. Hatsuda, Y. Miake, \textit{Quark-Gluons Plasma}%
, Cambridge University Press (2005).

\bibitem{IsospinR1} D.T. Son, M. A. Stephanov, Phys. Rev. Lett. 86 (2001)
592-595.

\bibitem{IsospinR2} J.B. Kogut, D.K. Sinclair, Phys. Rev. D 66 (2002) 034505.

\bibitem{IsospinR3} P.~Adhikari, J.~O.~Andersen and P.~Kneschke, 
%``Two-flavor chiral perturbation theory at nonzero isospin: Pion condensation at zero temperature,''
Eur. Phys. J. C \textbf{79}, no.10, 874 (2019).

\bibitem{IsospinR4} P.~Adhikari and J.~O.~Andersen, 
%``Quark and pion condensates at finite isospin density in chiral perturbation theory,''
Eur. Phys. J. C \textbf{80}, no.11, 1028 (2020).

\bibitem{IsospinR5} P. Adhikari, J. O. Andersen, M. A. Mojahed, Eur. Phys. J. C
81 (2021) 2, 173.

\bibitem{IsospinR6} J.~O.~Andersen, M.~Kj\o{}llesdal, Q.~Yu and H.~Zhou,
%``Chiral perturbation theory and Bose-Einstein condensation in QCD,''
[arXiv:2312.13092 [hep-ph]].

\bibitem{IsospinR7} T. Kojo, D. Suenaga, Phys. Rev. D 105 (2022) 7, 076001.

\bibitem{IsospinR8} N.~Kovensky and A.~Schmitt, 
%``Isospin asymmetry in holographic baryonic matter,''
SciPost Phys. \textbf{11}, no.2, 029 (2021).

\bibitem{IsospinR9} N.~Kovensky, A.~Poole and A.~Schmitt, 
%``Phases of cold holographic QCD: Baryons, pions and rho mesons,'' 
SciPost Phys. \textbf{15}, no.4, 162 (2023).

\bibitem{Pisarski2} R.~D.~Pisarski and F.~Wilczek, 
%``Remarks on the Chiral Phase Transition in Chromodynamics,''
Phys. Rev. D \textbf{29}, 338-341 (1984).

\bibitem{Pisarski3} L.~Y.~Glozman, O.~Philipsen and R.~D.~Pisarski, 
%``Chiral spin symmetry and the QCD phase diagram,''
Eur. Phys. J. A \textbf{58}, no.12, 247 (2022).

\bibitem{Brandt} B.~B.~Brandt, G.~Endrodi and S.~Schmalzbauer, 
%``QCD at finite isospin chemical potential,''
EPJ Web Conf. \textbf{175}, 07020 (2018).

\bibitem{Massimo2} S.~Carignano, L.~Lepori, A.~Mammarella, M.~Mannarelli and
G.~Pagliaroli, %``Scrutinizing the pion condensed phase,''
Eur. Phys. J. A \textbf{53}, no.2, 35 (2017).

\bibitem{Massimo3} M.~Mannarelli, %``Meson condensation,''
Particles \textbf{2}, no.3, 411-443 (2019).

\bibitem{[4]} N. Manton and P. Sutcliffe, Topological Solitons (Cambridge
University Press, Cambridge, 2007).

\bibitem{[5]} E. Shuryak, Nonperturbative Topological Phenomena in QCD and
Related Theories (Lecture Notes in Physics, 977, 2021 edition).

\bibitem{[6]} M. Shifman, Advanced Topics in Quantum Field Theory: A Lecture
Course, Cambridge University Press 2022.

\bibitem{WeinbergBook} E.~J.~Weinberg, 
``Classical solutions in quantum field theory: Solitons and Instantons in High Energy Physics,''
Cambridge University Press, 2012, ISBN 978-0-521-11463-9, 978-1-139-57461-7,
978-0-521-11463-9, 978-1-107-43805-7.

\bibitem{Adhikari1}
P.~Adhikari and J.~Choi,
%``Magnetic Vortices in the Abelian Higgs Model with Derivative Interactions,''
Int. J. Mod. Phys. A \textbf{33}, no.36, 1850215 (2019).

\bibitem{Adhikari2}
P.~Adhikari,
%``Magnetic Vortex Lattices in Finite Isospin Chiral Perturbation Theory,''
Phys. Lett. B \textbf{790}, 211-217 (2019).

\bibitem{Schmitt2} G.~W.~Evans and A.~Schmitt, 
%``Chiral anomaly induces superconducting baryon crystal,''
JHEP \textbf{09}, 192 (2022).

\bibitem{Schmitt3} G.~W.~Evans and A.~Schmitt, 
%``Chiral Soliton Lattice turns into 3D crystal,''
JHEP \textbf{2024}, no.02, 041 (2024).

\bibitem{Edery1}
A.~Edery,
%``Non-singular vortices with positive mass in 2+1 dimensional Einstein gravity with AdS$_3$ and Minkowski background,''
JHEP \textbf{01}, 166 (2021).

\bibitem{Edery2}
A.~Edery,
%``Nonminimally coupled gravitating vortex: Phase transition at critical coupling \ensuremath{\xi}c in AdS3,''
Phys. Rev. D \textbf{106}, no.6, 065017 (2022).

\bibitem{Nitta1} Z.~Qiu and M.~Nitta, 
%``Baryonic vortex phase and magnetic field generation in QCD with isospin and baryon chemical potentials,''
JHEP \textbf{06}, 139 (2024).

\bibitem{Nitta2} M.~Eto, K.~Nishimura and M.~Nitta, 
%``Phase diagram of QCD matter with magnetic field: domain-wall Skyrmion chain in chiral soliton lattice,''
JHEP \textbf{12}, 032 (2023).

\bibitem{Nitta3} M.~Eto, K.~Nishimura and M.~Nitta, 
%``How baryons appear in low-energy QCD: Domain-wall Skyrmion phase in strong magnetic fields,''
[arXiv:2304.02940 [hep-ph]].

\bibitem{Nitta4} M.~Eto and M.~Nitta, 
%``Chiral non-Abelian vortices and their confinement in three flavor dense QCD,''
Phys. Rev. D \textbf{104}, no.9, 094052 (2021).

\bibitem{US1} F. Canfora, \textit{Phys. Rev.} \textbf{D 88}, (2013), 065028; 
%``Ordered arrays of Baryonic tubes in the Skyrme model in ($3+1$) dimensions at finite density,''
\textit{Eur.\ Phys.\ J}.\ \textbf{C78}, no. 11, 929 (2018).

\bibitem{US2} F.~Canfora, S.~H.~Oh and A.~Vera, 
%``Analytic crystals of solitons in the four dimensional gauged non-linear sigma model,''
Eur. Phys. J. C \textbf{79}, no.6, 485 (2019).

\bibitem{US3} F.~Canfora, M.~Lagos and A.~Vera, 
%``Crystals of superconducting Baryonic tubes in the low energy limit of QCD at finite density,''
Eur. Phys. J. C \textbf{80}, no.8, 697 (2020).

\bibitem{US4} F. Canfora, S. Carignano, M. Lagos, M. Mannarelli, A. Vera,
Phys. Rev. D 103 (2021) 7, 076003.

\bibitem{US5} P. D. Alvarez, F. Canfora, N. Dimakis and A. Paliathanasis, 
\textit{Phys. Lett}. \textbf{B 773}, (2017) 401-407; F. Canfora, N. Dimakis,
A. Paliathanasis, \textit{Eur. Phys. J.} \textbf{C79} (2019) no.2, 139.

\bibitem{US6} S. L. Cacciatori, F. Canfora, M. Lagos, F. Muscolino, A. Vera,
JHEP 12 (2021) 150; Nucl. Phys. B 976 (2022) 115693. 
%doi:10.1140/epjc/s10052-018-6404-x
%[arXiv:1807.02090 [hep-th]].

\bibitem{Weinberg} E. J. Weinberg, Phys. Rev. D 19 (1979) 3008.

\bibitem{CPT1} S. Scherer, M. R. Schindler, \textit{A Primer for Chiral
Perturbation Theory}, Lecture Notes in Physics. Berlin Heidelberg:
Springer-Verlag (2012). ISBN 978-3-642-19253-1.

\bibitem{CPT2} J. Donoghue, E. Golowich, B. Holstein, \textit{Dynamics of
the Standard Model}, (Cambridge University Press, 1994).

\bibitem{CPT3} R. Machleidt, D. R. Entem, \textit{Physics Reports} \textbf{%
503} (1): 1--75 (2011).

\bibitem{1m} J. Gasser and H. Leutwyler, Ann. Phys., vol. 158, p. 142, 1984.

\bibitem{2m} H. Leutwyler, Ann. Phys., vol. 235, pp. 165\{203, 1994.

\bibitem{3m} G. Ecker, Prog. Part. Nucl. Phys., vol. \textbf{35}, 1-80, 1995.

\bibitem{4m} S. Scherer, Adv. Nucl. Phys., vol. 27, p. 277, 2003.

\bibitem{Massimo1} S.~Carignano, A.~Mammarella and M.~Mannarelli, 
%``Equation of state of imbalanced cold matter from chiral perturbation theory,''
Phys. Rev. D \textbf{93}, no.5, 051503 (2016).

\bibitem{BaMa} A. Balachandran, G. Marmo, B. Skagerstam, A. Stern, \textit{%
Classical Topology and Quantum States}, World Scientific (1991).

\bibitem{ex4d6} G. Baym, B. L. Friman and G. Grinstein, Nucl. Phys. B210
(1982) 193.

\bibitem{ex4d4} Y. Hidaka, K. Kamikado, T. Kanazawa and T. Noumi, Phys. Rev.
D92 (2015) 034003.

\bibitem{US7} F. Canfora, JHEP 11 (2023) 007.

\bibitem{strongB1} K.~Hattori, T.~Kojo and N.~Su, 
%``Mesons in strong magnetic fields: (I) General analyses,''
Nucl. Phys. A \textbf{951}, 1-30 (2016).

\bibitem{strongB2} K.~Hattori and X.~G.~Huang, 
%``Novel quantum phenomena induced by strong magnetic fields in heavy-ion collisions,''
Nucl. Sci. Tech. \textbf{28}, no.2, 26 (2017).

\bibitem{strongB3} H.~T.~Ding, S.~T.~Li, A.~Tomiya, X.~D.~Wang and Y.~Zhang, 
%``Chiral properties of (2+1)-flavor QCD in strong magnetic fields at zero temperature,''
Phys. Rev. D \textbf{104}, no.1, 014505 (2021).

\bibitem{US8} F. Canfora, H.~Maeda, \textit{Phys. Rev.} \textbf{D 87},
084049 (2013).

\bibitem{US9} F. Canfora, F. Correa, J. Zanelli, \textit{Phys. Rev.} \textbf{%
D 90}, 085002 (2014).

\bibitem{US10} F. Canfora, M. Di Mauro, M. A. Kurkov, A. Naddeo, \textit{%
Eur. Phys. J.} \textbf{C75} (2015) 9, 443.

\bibitem{US11} E. Ayon-Beato, F. Canfora, J. Zanelli, \textit{Phys. Lett.} 
\textbf{B 752, }(2016) 201-205; E.~Ay\'on-Beato, F.~Canfora, M.~Lagos,
J.~Oliva and A.~Vera, 
%``Analytic self-gravitating $4$-Baryons, traversable NUT-AdS wormholes, flat space-time multi-Skyrmions at finite volume and a novel transition in the $SU(3)$-Skyrme model,''
Eur. Phys. J. C \textbf{80}, no.5, 384 (2020).

\bibitem{US12} L. Aviles, F. Canfora, N. Dimakis, D. Hidalgo, \textit{Phys.
Rev. }\textbf{D 96} (2017), 125005; F.~Canfora, M.~Lagos, S.~H.~Oh, J.~Oliva
and A.~Vera, 
%``Analytic (3+1)-dimensional gauged Skyrmions, Heun, and Whittaker-Hill equations and resurgence,''
Phys.\ Rev.\ D \textbf{98}, no. 8, 085003 (2018). 
%doi:10.1103/PhysRevD.98.085003
%[arXiv:1809.10386 [hep-th]].

\bibitem{US13} F.~Canfora, D.~Hidalgo, M.~Lagos, E.~Meneses and A.~Vera, 
%``Infinite conformal symmetry and emergent chiral fields of topologically nontrivial configurations: From Yang-Mills-Higgs theory to the Skyrme model,''
Phys. Rev. D \textbf{106}, no.10, 105016 (2022).

\bibitem{US14} S. L. Cacciatori, F. Canfora, F. Muscolino, Nucl. Phys. B 1000
(2024) 116477.

\bibitem{Barducci} A.~Barducci, G.~Pettini, L.~Ravagli and R.~Casalbuoni, 
%``Ladder QCD at finite isospin chemical potential,''
Phys. Lett. B \textbf{564}, 217-224 (2003)

\bibitem{Loewe} M.~Loewe, S.~Mendizabal and J.~C.~Rojas, 
%``Skyrme model and isospin chemical potential,''
Phys. Lett. B \textbf{632}, 512-516 (2006).

\bibitem{Scoccola} J.~A.~Ponciano and N.~N.~Scoccola, 
%``Skyrmions in the presence of isospin chemical potential,''
Phys. Lett. B \textbf{659}, 551-554 (2008).

\bibitem{Cao} X.~Cao, H.~Liu, D.~Li and G.~Ou, 
%``QCD phase diagram at finite isospin chemical potential and temperature in an IR-improved soft-wall AdS/QCD model,''
Chin. Phys. C \textbf{44}, no.8, 083106 (2020).

\bibitem{YsonnYang} S. Wang, Y. Yang, \textit{Siam Journal of Mathematical
Analysis} \textbf{23} (1992), 1125-1140.

\bibitem{Taubes1} C. H. Taubes, \textit{Comm. Math. Phys}. \textbf{72},
277-292 (1980).

\bibitem{Taubes2} A. Jaffe, C. H. Taubes, \textit{Vortices and Monopoles},
Birkhauser, Boston (1980).

\bibitem{WeinbergIndex}
E. J. Weinberg,
%``Index Calculations for the Fermion-Vortex System,''
Phys. Rev. D \textbf{24}, 2669 (1981).

\bibitem{Marino}
M. Mariño,  Instantons and Large N: An Introduction to Non-Perturbative Methods in Quantum Field Theory,
Cambridge University Press, 2015.
ISBN 978-1-107-06852-0, 978-1-316-37154-1.

\bibitem{Roadmap}
Lilia Boeri et al 2022 J. Phys.: Condens. Matter \textbf{34} 183002.


\bibitem{Jackiw:1981ee}
R.~Jackiw and P.~Rossi,
Nucl. Phys. B \textbf{190} (1981), 681-691.

\bibitem{Balachandran:1982cb}
A.~P.~Balachandran, V.~P.~Nair, S.~G.~Rajeev and A.~Stern,
%``Soliton States in the QCD Effective Lagrangian,''
Phys. Rev. D \textbf{27} (1983), 1153
[erratum: Phys. Rev. D \textbf{27} (1983), 2772].

\bibitem{Balachandran:1982dw}
A.~P.~Balachandran, V.~P.~Nair, S.~G.~Rajeev and A.~Stern,
%``Exotic Levels from Topology in the QCD Effective Lagrangian,''
Phys. Rev. Lett. \textbf{49} (1982), 1124
[erratum: Phys. Rev. Lett. \textbf{50} (1983), 1630].

\bibitem{Kahana:1985ycl}
S.~Kahana, R.~Perry and G.~Ripka,
%``INFINITE COUPLING LIMIT OF A CHIRAL FIELD,''
Phys. Lett. B \textbf{163} (1985), 37-40.

\bibitem{mass1} S.~B.~Gudnason and M.~Nitta, 
%``Modifying the pion mass in the loosely bound Skyrme model,''
Phys. Rev. D \textbf{94}, no.6, 065018 (2016).

\bibitem{mass2} S.~B.~Gudnason, %``Loosening up the Skyrme model,''
Phys. Rev. D \textbf{93}, no.6, 065048 (2016).

\bibitem{mass3} G.~S.~Adkins and C.~R.~Nappi, 
%``The Skyrme Model with Pion Masses,''
Nucl. Phys. B \textbf{233}, 109-115 (1984).

\bibitem{mass4} V.~B.~Kopeliovich, B.~Piette and W.~J.~Zakrzewski, 
%``Mass terms in the Skyrme model,''
Phys. Rev. D \textbf{73}, 014006 (2006).

\bibitem{mass5} L.~Marleau, 
%``Modifying the Skyrme model: Pion mass and higher derivatives,''
Phys. Rev. D \textbf{43}, 885-890 (1991).

\bibitem{Dorso} J. A. Lopez, C. O. Dorso, G. A. Frank, Front.Phys. (Beijing)
16 (2021) 2, 24301.

\bibitem{pasta2a} C. J. Horowitz, D. K. Berry, C.M. Briggs, M. E. Caplan, A.
Cumming, A. S. Schneider, Phys. Rev. Lett. 114, 031102 (2015).

\bibitem{pasta2b} D. K. Berry, M. E. Caplan, C. J. Horowitz, G. Huber, A. S.
Schneider, Phys. Rev. C 94, 055801 (2016).

\bibitem{pasta3} C. O. Dorso, G. A. Frank, J. A. L\'{o}pez, Nucl. Phys.
A978, 35 (2018).

\bibitem{pasta4} A. da Silva Schneider, M. E. Caplan, D. K. Berry, C. J.
Horowitz, Phys. Rev. C 98, 055801 (2018).

\bibitem{pasta5} M. E. Caplan, A. S. Schneider, and C. J. Horowitz, Phys.
Rev. Lett. 121, 132701 (2018).

\bibitem{pasta6} R. Nandi and S. Schramm, J. Astrophys. Astron. 39, 40
(2018).

\bibitem{pasta7} Z. Lin, M. E. Caplan, C. J. Horowitz, C. Lunardini, Phys.
Rev. C 102 (2020) 4, 045801.

\bibitem{pasta8} C.O. Dorso, A. Strachan, G.A. Frank, Nucl. Phys. A 1002
(2020) 122004.

\bibitem{pasta9} C.J. Pethick, Z. Zhang, D.N. Kobyakov, Phys. Rev. C 101
(2020) 5, 055802.

\bibitem{Watanabe} G.~Watanabe and T.~Maruyama, 
%``Nuclear pasta in supernovae and neutron stars,''
[arXiv:1109.3511 [nucl-th]].

\bibitem{Schmitt1} A.~Schmitt, 
%``Chiral pasta: Mixed phases at the chiral phase transition,''
Phys. Rev. D \textbf{101}, no.7, 074007 (2020).

\bibitem{Manton1} N.S. Manton, Phys. Lett. B 110 (1982) 54; Phys. Lett. B
154 (1985) 397; (E) B 157 (1985) 475.

\bibitem{Manton2} N.S. Manton, S.M. Nasir, Commun.Math.Phys. 199 (1999)
591-604.

\bibitem{Manton3} G.W. Gibbons, N.S. Manton, Phys.Lett.B 356 (1995) 32-38.

\bibitem{Samols} T.M. Samols, \textit{Commun.Math.Phys.} \textbf{145} (1992)
149-180.

\bibitem{GilbargTrudinger} D. Gilbarg, N. Trudinger, \textit{Elliptic
Partial Differential Equations of Second Order}, Sprimger (2013).

\bibitem{BanachCaccioppoli1} D. Kinderlehrer, G. Stampacchia, \textit{%
Variational Inequalities in} $R^{N}$: \textit{An Introduction to Variational
Inequalities and Their Applications,} New York: Academic Press (1980), pp.
7--22.

\bibitem{BanachCaccioppoli2} P. Pedregal, \textit{Functional Analysis,
Sobolev Spaces, and Calculus of Variations}, Springer 2024.

\end{thebibliography}
\end{document}